\documentclass[aps,prc,twocolumn,superscriptaddress,floatfix,amsmath,amssymb,nofootinbib,longbibliography,colorlinks=true,citecolor=blue,linkcolor=blue,urlcolor=blue]{revtex4-2}

\usepackage{graphicx}
\usepackage{dcolumn}
\usepackage{bm}
\usepackage{orcidlink}
\usepackage{hyperref}

\newcommand{\cgz}[3]{\mathcal{C}_{#1 0 #2 0}^{#3 0}}
\newcommand{\cg}[6]{\mathcal{C}_{#1 #4 #2 #5}^{#3 #6}}

\newcommand{\sixj}[6]{\left\{
    \begin{array}{ccc}
      #1 & #2 & #3 \\
      #4 & #5 & #6
    \end{array}
\right\}}
\newcommand{\ninej}[9]{\left\{
    \begin{array}{ccc}
      #1 & #2 & #3 \\
      #4 & #5 & #6 \\
      #7 & #8 & #9
    \end{array}
\right\}}
\renewcommand{\vec}[1]{\boldsymbol{#1}}
\newcommand{\uvec}[1]{\widehat{\boldsymbol{#1}}}

\begin{document}

\title{Two-body currents at finite momentum transfer\\
and applications to M1 transitions}

\author{C. Brase\,\orcidlink{0000-0002-5876-7621}}
\email{catharina.brase@tu-darmstadt.de}
\affiliation{Technische Universit\"at Darmstadt, Department of Physics, 64289 Darmstadt, Germany}
\affiliation{ExtreMe Matter Institute EMMI, GSI Helmholtzzentrum f\"ur Schwerionenforschung GmbH, 64291 Darmstadt, Germany}
\affiliation{Max-Planck-Institut f\"ur Kernphysik, Saupfercheckweg 1, 69117 Heidelberg, Germany}

\author{T.~Miyagi\,\orcidlink{0000-0002-6529-4164}}
\email{miyagi@nucl.ph.tsukuba.ac.jp}
\affiliation{Center for Computational Sciences, University of Tsukuba, 1-1-1 Tennodai, Tsukuba 305-8577, Japan}
\affiliation{Technische Universit\"at Darmstadt, Department of Physics, 64289 Darmstadt, Germany}
\affiliation{ExtreMe Matter Institute EMMI, GSI Helmholtzzentrum f\"ur Schwerionenforschung GmbH, 64291 Darmstadt, Germany}
\affiliation{Max-Planck-Institut f\"ur Kernphysik, Saupfercheckweg 1, 69117 Heidelberg, Germany}

\author{J. Men\'endez\,\orcidlink{0000-0002-1355-4147}}
\email{menendez@fqa.ub.edu}
\affiliation{Departament de F\'isica Qu\`antica i Astrof\'isica, Universitat de Barcelona, 08028 Barcelona, Spain}
\affiliation{Institut de Ci\`encies del Cosmos, Universitat de Barcelona, 08028 Barcelona, Spain}

\author{A.~Schwenk\,\orcidlink{0000-0001-8027-4076}}
\email{schwenk@physik.tu-darmstadt.de}
\affiliation{Technische Universit\"at Darmstadt, Department of Physics, 64289 Darmstadt, Germany}
\affiliation{ExtreMe Matter Institute EMMI, GSI Helmholtzzentrum f\"ur Schwerionenforschung GmbH, 64291 Darmstadt, Germany}
\affiliation{Max-Planck-Institut f\"ur Kernphysik, Saupfercheckweg 1, 69117 Heidelberg, Germany}

\begin{abstract}
We explore the impact of two-body currents (2BCs) at finite momentum transfer with a focus on magnetic dipole properties in $^{48}$Ca and $^{48}$Ti. To this end, we derive a multipole decomposition of 2BCs to fully include the momentum-transfer dependence in {\it ab initio} calculations. As application, we investigate the effects of 2BCs on the strong M1 transition at 10.23\,MeV in $^{48}$Ca using the valence-space in-medium similarity renormalization group (VS-IMSRG) with a set of non-implausible interactions as well as the 1.8/2.0 (EM) interaction. Experiments, such as $(e,e')$ and $(\gamma,n)$, disagree on the magnetic dipole strength $B$(M1) for this transition. Our VS-IMSRG results favor larger $B$(M1) values similar to recent coupled-cluster calculations. However, for this transition there are larger cancellations between the leading pion-in-flight and seagull 2BCs, so that future calculations including higher-order 2BCs are important. For validation of our results, we investigate additional observables in $^{48}$Ca as well as M1 transitions in $^{48}$Ti. For these, our results agree with experiment. Finally, our results show that for medium-mass nuclei 2BC contributions to M1 and Gamow-Teller transitions are, as expected, very different. Therefore, using similar quenching factors for both in phenomenological calculations is not supported from first principles.
\end{abstract}

\maketitle

\section{Introduction}\label{sec:intro}
Nuclear electroweak processes can be described in a first approximation using one-body operators that involve a single nucleon. However, over a decade ago, few-body calculations established that two-body currents (2BCs) involving the coupling to two nucleons play a significant role in the weak~\cite{Nakamura:2000vp,Butler2001,Gazit2009} and electromagnetic~\cite{Bacca:2014tla,Marcucci:2015rca} sectors. Recent {\it ab initio} calculations of electromagnetic moments and form factors~\cite{Lynn:2019rdt,Gnech:2022vwr,Seutin:2023grs,Martin:2023dhl,Pal:2023gll,Chambers-Wall2024}, $\beta$-decay lifetimes and spectra~\cite{Gysbers2019,King2020,King:2022zkz}, muon capture~\cite{King2022} as well as electron-nucleus~\cite{Lynn:2019rdt,Andreoli:2024ovl} and neutrino-nucleus~\cite{Lynn:2019rdt,Lovato:2020kba} scattering, highlight the importance of 2BCs.

In heavier systems the inclusion of 2BCs is more challenging. A complete evaluation has been performed in $\beta$ decays~\cite{Gysbers2019},  where it is a good approximation to include 2BCs at vanishing momentum transfer, and very recently in magnetic-dipole (M1) transitions~\cite{Miyagi2024,Acharya2024}. However, for other processes involving medium-mass and heavy nuclei, calculations are based on approximations for the 2BCs at finite momentum transfer due to the complexities involved in deriving 2BC matrix elements and other computational limitations. This is the case for muon capture \cite{Jokiniemi:2021qqg,Gimeno2023,Jokiniemi2024}, neutrino-nucleus scattering \cite{Hoferichter2020}, neutrinoless $\beta\beta$ decays~\cite{Menendez2011,Engel2014,Jokiniemi:2022ayc} or the scattering of weakly-interacting massive particles off nuclei~\cite{Menendez2012a,Klos2013a,Baudis2013,Hoferichter:2016nvd,Hoferichter2019}. For the latter two processes, the relevant nuclear matrix elements are needed to extract information about physics beyond the standard model of particle physics from experimental searches~\cite{Engel2017a,Aalbers:2022dzr}. 

In order to calculate electroweak processes without approximating 2BCs at finite momentum transfer, a multipole decomposition of the currents is needed. In this work, we present a derivation of the multipole decomposition of 2BCs at finite momentum transfer. We implement the multipole-decomposed 2BCs in {\it ab initio} valence space in-medium similarity renormalization group (VS-IMSRG) calculations, which describe well the nuclear structure and electroweak observables of medium-mass and heavy nuclei~\cite{Stroberg:2016ung,Morris:2017vxi,Stroberg:2019mxo,Stroberg:2019bch,Gysbers2019,Miyagi:2021pdc,Hebeler:2022aui,Heinz:2024juw}.

As a first application of this framework, we study the strong M1 transition in $^{48}$Ca~\cite{Steffen1980} from the $0^+$ ground state to the excited $1^+$ state at $10.23$\,MeV. This transition contains most of the $B$(M1) transition strength in this nucleus, and the transition form factor is also the best measured M1 transition form factor~\cite{Steffen1983}. We present {\it ab initio} VS-IMSRG calculations of the $B$(M1) strength and the transition form factor.

On the experimental side, experiments based on $(e,e')$ scattering~\cite{Steffen1980,Steffen1983} and $(\gamma,n)$ reactions~\cite{Tompkins2011} disagree on the value of this $B$(M1) strength, with the latter measurements pointing to a value almost twice as large as the $(e,e')$ experiments. More recently, a $(p,p')$ experiment has favored the smaller $B$(M1) strength value found in the $(e,e')$ experiments~\cite{Birkhan2016}. Because of this tension, we also explore the $B$(M1) strength in the neighboring nucleus $^{48}$Ti, which is also known experimentally~\cite{Guhr1990}. Moreover, we estimate our theoretical uncertainties by considering a set of 34 non-implausible interactions~\cite{Hu2022} and the 1.8/2.0 (EM) Hamiltonian~\cite{Hebeler2011}.

Very recently, this $B$(M1) transition in $^{48}$Ca has also been studied with {\it ab initio} coupled-cluster calculations~\cite{Acharya2024}. This work used four different Hamiltonians and included continuum effects at the one-body-current (1BC) level. The coupled-cluster results favor larger $B$(M1) values in agreement with the $(\gamma,n)$ experiment, with a small effect due to 2BCs. In this work, we revisit these findings with the VS-IMSRG approach with a 2BC multipole decomposition, which enables calculations of the transition form factor for a wide range of momentum transfers. We also analyze the different contributions from 2BCs. Our results indicate that future calculations including higher-order 2BCs are important, because for this transition there are larger cancellations between different contributions to 2BCs. Finally, we also compare the role of 2BCs for M1 and Gamow-Teller (GT) transitions, and show that the vector and axial-vector  2BC effects are very different.

This paper is structured as follows. In Sec.~\ref{sec:many_body} we discuss the many-body method and applied chiral interactions. In Sec.~\ref{sec:MD_2BCs} we present the multipole decomposition of the leading electromagnetic 2BCs, followed by their application in Sec.~\ref{sec:TFF} to the transition form factor of the most dominant M1 transition in $^{48}$Ca. In Secs.~\ref{sec:M1_48Ca} and~\ref{sec:M1_48Ti} we present our calculations of the $B$(M1) strengths in $^{48}$Ca and $^{48}$Ti, respectively. Finally, we conclude and give an outlook in Sec.~\ref{sec:conclusion}.

\section{Many-body calculation}\label{sec:many_body}
In this work, we use as many-body method the {\it ab initio} VS-IMSRG~\cite{Tsukiyama:2012sm,Stroberg:2016ung,Stroberg:2019mxo}, which decouples a valence space (in this work the $pf$ shell) from excitations outside the valence space. The VS-IMSRG starts from the intrinsic Hamiltonian
\begin{equation}
    H = T - T_\mathrm{CM} + V_\mathrm{NN} + V_\mathrm{3N}\,,
\end{equation}
where $T - T_\mathrm{CM}$ denotes the kinetic energy of the $A$ nucleons subtracted by the kinetic energy of the center-of-mass (CM) motion. $V_\mathrm{NN}$ and $V_\mathrm{3N}$ are the two-nucleon (NN) and three-nucleon (3N) interactions.

To explore the uncertainties from the Hamiltonian parameters, we consider the set of 34 non-implausible NN+3N interactions~\cite{Hu2022} based on $\Delta$-full chiral EFT, as well as the 1.8/2.0 (EM) NN+3N interaction~\cite{Hebeler2011}. The non-implausible interactions are sampled based on the implausibility measure defined by low-energy NN phase shifts, properties of $A=2,3,4$ nuclei, as well as information from energies and radii of $^{16}$O~\cite{Hu2022}. The 1.8/2.0 (EM) Hamiltonian reproduces NN scattering for $E_\mathrm{lab} \lesssim 300$\,MeV, and the 3N couplings are fit to the $^3$H energy and the $^4$He radius~\cite{Hebeler2011}.

We truncate the VS-IMSRG evolution at the normal-ordered two-body level, yielding the VS-IMSRG(2) approximation. The M1 operator is consistently evolved, both at the 1BC and 2BC level in the VS-IMSRG using the Magnus formulation~\cite{Morris:2015yna} and keeping all operator contributions at the normal-ordered two-body level, as was done for the magnetic moment calculations in Ref.~\cite{Miyagi2024}. For the calculations of the interaction and current matrix elements we use the \texttt{NuHamil} code~\cite{Miyagi2023}, for the VS-IMSRG calculations the \texttt{IMSRG++} code~\cite{Stroberg2018}, and for the valence-space diagonalization and transition density calculations the \text{KSHELL} code~\cite{Shimizu:2019xcd}.

\begin{figure}[t!]
    \centering
    \includegraphics[width=1.0\linewidth]{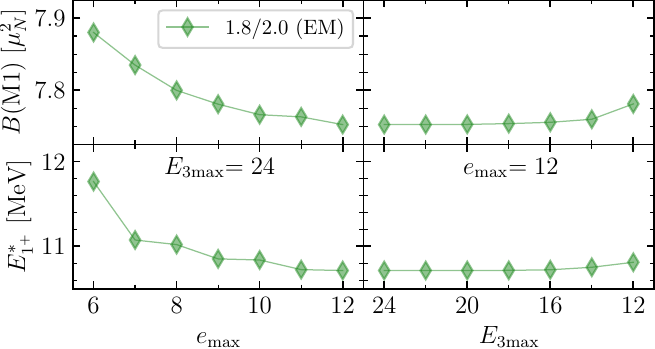}
    \caption{Many-body convergence with respect to the basis parameters $e_\mathrm{max}$ and $E_\mathrm{3max}$ for the M1 strength (top) and the $1^+$ excitation energy (bottom) of $^{48}$Ca using the VS-IMSRG for the 1.8/2.0 (EM) interaction. The left (right) panels show the variation with $e_\mathrm{max}$ ($E_\mathrm{3max}$) for $E_\mathrm{3max}=24$ ($e_\mathrm{max}=12$).}
    \label{fig:conv}
\end{figure}

All our calculations use a harmonic-oscillator frequency $\hbar \omega = 16$\,MeV and are performed in the Hartree-Fock basis. To check the many-body convergence of the VS-IMSRG, we have studied the M1 strength, $B$(M1), and the $1^+$ excitation energy, $E^*_{1^+}$, of $^{48}$Ca as a function of the basis parameters $e_\mathrm{max}={\rm max}(2n+l)$ and $E_\mathrm{3max}={\rm max}(2n_1+l_1+2n_2+l_2+2n_3+l_3)$, where $n$ and $l$ denote the principal and angular momentum quantum numbers of the single-particle basis states. Figure~\ref{fig:conv} shows that the M1 strength and the excitation energy are well converged with respect to $E_\mathrm{3max}$ at $E_\mathrm{3max}=24$, while there is still a small change going from $e_\mathrm{max}=10$ to 12. Based on this, the $B$(M1) and $E^*_{1^+}$ are converged (with respect to the many-body basis) at percent level or better. In the following, we perform VS-IMSRG(2) calculations with $e_\mathrm{max}=12$ and $E_\mathrm{3max}=24$.

\section{Multipole decomposition of two-body currents}\label{sec:MD_2BCs}
To obtain further insight, we compute the magnetic transition form factor as well.
To do so, one would need a multipole decomposition of the current.
For the 1BC, the expression is well known.
However, the multipole decomposition of the leading 2BCs is less known.
The matrix element of the 2BC is known in the plane-wave basis, and we need to define the momenta: incoming momentum $\vec{p}_{1}$ ($\vec{p}_{2}$) of nucleon 1 (2), and outgoing momentum $\vec{p}_{1}'$ ($\vec{p}_{2}'$).
The momentum carried by the photon is defined as $\vec{Q} = \vec{p}'_{1} + \vec{p}'_{2} - \vec{p}_{1} - \vec{p}_{2}$.
Also, we define the momentum transfer $\vec{q}_{i} = \vec{p}'_{i} - \vec{p}_{i}$ $(i=1,2)$.
Introducing the two-body plane-wave basis $|\vec{p}_{1}s_{1}, \vec{p}_{2}s_{2} \rangle$ with $z$-component of the spin $s_{i}$ ($i=1,2$), the leading electromagnetic 2BCs are given by~\cite{Krebs2019,Krebs2020} (see also Ref.~\cite{Seutin:2023grs} for our conventions):
\begin{equation}
\begin{aligned} 
\label{eq:2BC}
 \langle \vec{p}_{1}'s_{1}', &\vec{p}_{2}'s_{2}'  | \vec{j}(\vec{Q})| \vec{p}_{1}s_{1}, \vec{p}_{2}s_{2} \rangle = -ie \frac{g_{A}^{2}}{4F_{\pi}^{2}}  (\vec{\tau}_1 \times \vec{\tau}_2)_{z}  
 \\ & \times
 \left[ \vec{\sigma}_1 - \vec{q}_{1} \frac{\vec{\sigma}_1 \cdot \vec{q}_1}{q_1^2 + m_\pi^2} \right] 
\frac{\vec{\sigma}_2 \cdot \vec{q}_2}{q_2^2 + m_\pi^2} + 1 \leftrightharpoons 2 \,,
\end{aligned}
\end{equation}
with the elementary charge $e$, the axial coupling $g_A = 1.27$, the pion decay constant $F_\pi = 92.4$\,MeV, and the averaged pion mass $m_\pi = 138.039$\,MeV.

To achieve a numerical calculation with a spherically formulated method, one needs matrix elements of multipole components of $\vec{j}(\vec{Q})$ expressed in the basis $|ab : J_{ab}\rangle$, where the single-particle states $a$ and $b$ are coupled to total angular momentum $J_{ab}$.
Here, the single-particle orbital $a$ is specified with the harmonic-oscillator quantum numbers: radial quantum number $n_{a}$, orbital angular momentum $l_{a}$, total angular momentum $j_{a}$, $z$-component of the total angular momentum $m_{a}$, and $z$-component of the isospin $\tau_{z,a}$ for proton or neutron.
The multipole decomposition of $\vec{j}(\vec{Q})$ can be written as
\begin{widetext}
\begin{equation}
\vec{j}(\vec{Q}) = 4\pi \sum_{\lambda\mu} i^{\lambda+1} \left[
L_{\lambda\mu}(Q)\vec{Y}^{*}_{\lambda\mu}\bigl(\uvec{Q}\bigr) 
+ T^{\rm el}_{\lambda\mu}(Q)\vec{\Psi}^{*}_{\lambda\mu}\bigl(\uvec{Q}\bigr) 
+ T^{\rm mag}_{\lambda\mu}(Q)\left[i\vec{\Phi}_{\lambda\mu}\bigl(\uvec{Q}\bigr)\right]^{*}
\right],
\end{equation}
with vector spherical harmonics $\vec{Y}_{\lambda\mu}(\uvec{Q})$, $\vec{\Psi}_{\lambda\mu}(\uvec{Q})$, and $\vec{\Phi}_{\lambda\mu}(\uvec{Q})$.
Note that $\uvec{x}$ indicates the direction of $\vec{x}$.
The definitions of the vector spherical harmonics are given in Appendix~\ref{sec:multipoles}.
The above decomposition manifests that $L_{\lambda\mu}(Q)$, $T^{\rm el}_{\lambda\mu}(Q)$, and $T^{\rm mag}_{\lambda\mu}(Q)$ are spherical tensors.
To compute a magnetic form factor, one needs the reduced matrix element $\langle a'b':J_{a'b'}|| T^{\rm mag}_{\lambda} || ab :J_{ab}\rangle$.

A direct way to compute the matrix element from Eq.~\eqref{eq:2BC} would be to integrate with respect to $\vec{p}_{1}$, $\vec{p}_{2}$, $\vec{p}_{1}'$, and $\vec{p}_{2}'$ through
\begin{equation}
\begin{aligned}
\langle \vec{p} s | a \rangle &= (-i)^{l_{a}} 
R_{n_{a}l_{a}}(p) \, Y_{l_{a},m_{l}}(\uvec{p}) \, \cg{l_{a}}{\frac{1}{2}}{j_{a}}{m_{l}}{s}{m_{a}}\,, \\
R_{n_{a}l_{a}}(p) &= (-1)^{n_{a}} 
\sqrt{\frac{2b_{\rm osc}^{3}\Gamma(n_{a}+1)}{\Gamma(n_{a}+l_{a}+3/2)}}  (pb_{\rm osc})^{l_{a}} L^{(l_{a}+1/2)}_{n_{a}}(p^{2}b^{2}_{\rm osc}) \, e^{-p^{2}b^{2}_{\rm osc}/2}\,.
\end{aligned}
\end{equation}
Here, $p$ is the magnitude of $\vec{p}$, $b_{\rm osc}$ is the oscillator length of the radial basis function, $\Gamma(x)$ is the Gamma function, $L^{\alpha}_{n}(x)$ are associated Laguerre polynomials, $Y_{l,m_{l}}(\uvec{x})$ are spherical harmonics, and $\cg{j_{1}}{j_{2}}{J}{m_{1}}{m_{2}}{M}$ are Clebsch-Gordan coefficients.
However, the direct integral procedure is numerically too expensive as it requires 11-dimensional integrals.
To make the computation feasible, we can first calculate the matrix element in the partial-wave basis with the relative and center-of-mass motions $|Pp\alpha \rangle$, where $P$ and $p$ are the magnitudes of center-of-mass and relative momenta, and $\alpha$ is a collective index for the angular momentum and spin quantum numbers.
The index $\alpha$ is defined as $\alpha = (L, l, L_{\rm tot}, S, J)$ with the orbital angular momentum of the center-of-mass motion $L$, the orbital angular momentum of the relative motion $l$, total orbital angular momentum $\vec{L}_{\rm tot} = \vec{L}+\vec{l}$, total spin $S$, and total angular momentum $\vec{J} = \vec{L}_{\rm tot} + \vec{S}$.
With this basis, the matrix element can be computed as
\begin{equation}
\langle P' p' \alpha' || T^{\rm mag}_{\lambda}(Q) || Pp\alpha \rangle = -\frac{i^{L'+l'-L-l-\lambda-1}eg_{A}^{2}}{16 \pi F^{2}_{\pi} } i(\vec{\tau}_{1} \times \vec{\tau}_{2})_{z}
\left[\mathcal{A}^{\lambda,\lambda}_{\alpha'\alpha}(P',p',P,p,Q)+2\mathcal{B}^{\lambda,\lambda}_{\alpha'\alpha}(P',p',P,p,Q)\right].
\end{equation}
The functions $\mathcal{A}^{\lambda,\lambda}_{\alpha'\alpha}(P',p',P,p,Q)$ and  $\mathcal{B}^{\lambda,\lambda}_{\alpha'\alpha}(P',p',P,p,Q)$ are rather complicated objects and are given in Eqs.~\eqref{eq:functionA} and~\eqref{eq:functionB}, respectively, in the Appendix~\ref{sec:multipoles}, where the matrix elements of the other multipole operators $L_{\lambda\mu}(Q)$ and $T^{\rm el}_{\lambda\mu}(Q)$ are also discussed.
Then, one can obtain the matrix element expressed in the center-of-mass and relative harmonic-oscillator basis $|Nn\alpha \rangle$:
\begin{equation}
\label{eq:MomToHO}
\begin{aligned}
\langle N'n' \alpha' || T^{\rm mag}_{\lambda\mu} || Nn\alpha \rangle =
\int dP' dp' dP dp P'^{2}p'^{2}P^{2}p^{2}
R_{N'L'}\left(P'\right)
R_{n'l'}\left(p'\right) &
R_{NL}\left(P\right)
R_{nl}\left(p\right) 
\\ & \times
\langle P'p'\alpha' || T^{\rm mag}_{\lambda} || Pp \alpha \rangle \,.
\end{aligned}
\end{equation}
Finally, the matrix element
$\langle a'b' : J_{a'b'} || T^{\rm mag}_{\lambda} || ab: J_{ab} \rangle$ can be obtained through the Talmi-Moshinsky transformation~\cite{Kamuntavicius2001}:
\begin{equation}
\label{eq:TMTrans}
\langle a'b' : J_{a'b'} || T^{\rm mag}_{\lambda} || ab: J_{ab} \rangle = 
\sum_{N'n'\alpha'} \sum_{Nn\alpha} 
\langle a'b' : J_{a'b'} | N'n' \alpha' \rangle 
\langle N'n' \alpha' || T^{\rm mag}_{\lambda} || Nn\alpha \rangle
\langle Nn\alpha | ab: J_{ab} \rangle \,.
\end{equation}
\end{widetext}
The expression for the overlap $\langle Nn\alpha | ab : J_{ab} \rangle$ is given in Eq.~\eqref{eq:TMbracket} in Appendix~\ref{sec:multipoles}.
The actual implementation can be found in the \texttt{NuHamil} code~\cite{Miyagi2023}.

\section{Finite momentum transfer and transition form factor in $^{48}$Ca}\label{sec:TFF}
The transition form factor $F_\mathrm{T}(Q^2)$ from the $0^+$ ground state to the $1^+$ excited state at $10.23$\,MeV in $^{48}$Ca is calculated from the transverse magnetic multipole of the vector current operator for multipolarity 1,
\begin{equation}
    \begin{aligned}
        F^2_\mathrm{T}(Q^2) = \frac{4\pi}{Z^2} |\langle 1^+||T^\mathrm{mag}_{1}(Q)||0^+ \rangle|^2\,,
    \end{aligned}
\end{equation}
where $Z$ is the proton number. To theoretically identify the $1^{+}$ state corresponding to the experimental $10.23$\,MeV state, we computed the 10 lowest $1^{+}$ states and clearly observed the state showing the dominant $B$(M1) connected to the ground state. For 25 (33) out of the 34 interactions, the second largest $B$(M1) is less than $1\%$ ($10\%$) of the largest one. There is only one exception, showing a non-negligible second largest $B$(M1), where it is about $53\%$ of the largest one. Note that the second largest $B$(M1) is also less than $1\%$ of the largest one for the 1.8/2.0~(EM) interaction.

\begin{figure}[t!]
    \centering
    \includegraphics[width=\linewidth]{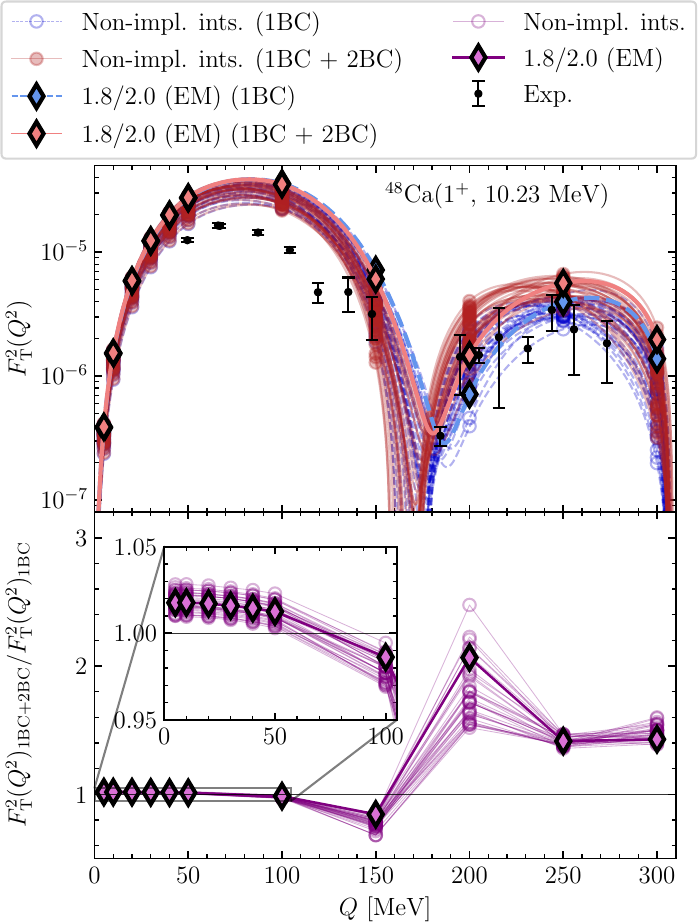}
    \caption{Top panel: Transition form factor $F^2_\mathrm{T}(Q^2)$ as a function of the momentum transfer $Q$. Results are shown for the non-implausible interactions with only 1BCs (open blue circles) and including also 2BCs (1BCs+2BCs, filled red circles). The blue (red) filled diamonds show the result for the 1.8/2.0 (EM) interaction. In order to better compare the VS-IMSRG results to the $(e,e')$ experimental data \cite{Steffen1983} (black points with error bars), we interpolate our results using \texttt{scipy.interpolate.CubicSpline} \cite{Virtanen2020}. Bottom panel: Ratio between the transition form factor with 2BCs and without, $F^2_\mathrm{T}(Q^2)_\mathrm{1BC+2BC}/F^2_\mathrm{T}(Q^2)_\mathrm{1BC }$ as function of momentum transfer $Q$ for the same interactions as in the top panel.}
    \label{fig:TFF}
\end{figure}

Our VS-IMSRG results for the transition form factor with and without 2BCs are shown in Fig.~\ref{fig:TFF} for the non-implausible interactions and the 1.8/2.0 (EM) interaction in comparison to the experimental data from inelastic electron scattering \cite{Steffen1983}. We find that the overall trend is similar to experiment. In more detail, up to $Q \lesssim 100$\,MeV, there is only a small contribution from 2BCs which increases the transition form factor in the low-$Q$ limit, see also the lower panel of Fig.~\ref{fig:TFF}. However, within the spread from the different Hamiltonians, our results overestimate the experimental data. At higher momentum transfer, there is overlap between experiment and the theoretical calculations, but there is also a wider spread in the predictions and a bigger experimental uncertainty.

To understand the impact of 2BCs on the transition form factor in more detail, the bottom panel of Fig.~\ref{fig:TFF} shows the relative contribution from 2BCs. Note that since the quantity is the square of the transition form factor, there is also a cross term between 1BCs and 2BCs. As expected from the momentum transfer dependence of the 2BCs, the relative importance of 2BCs increases with increasing $Q$. Moreover, at larger $Q$ the spread from the Hamiltonian dependence is larger. Up to $Q \lesssim 100$\,MeV, the combined contribution from 2BCs and their interference with the 1BCs is about two orders of magnitude smaller than the leading-order 1BCs only, leading to an increase of the transition form factor below 50\,MeV of up to $3\%$ only, for all interactions considered. As we discuss later, this small effect is also due to cancellations between the different parts of the 2BCs. This is different at higher momentum transfer. Here, the relative impact of 2BCs varies more with momentum transfer and for different interactions.

In the following section, we study the M1 strength $B$(M1) of the same transition. This strength can be obtained directly via the dipole operator or using the $Q \rightarrow 0$ limit of the transition form factor (see, e.g., Ref.~\cite{Acharya2024}): 
\begin{equation}
\lim_{Q \rightarrow 0}\frac{4\pi}{Z^2Q^2}|\langle 1^+||T^\mathrm{mag}_{1}(Q)||0^+ \rangle|^2= \frac{8\pi}{9Z^2} B(\mathrm{M1}; 0^{+}\to1^{+}) \,.
\end{equation}
We have checked that the M1 strength calculated from the dipole operator and from the low-$Q$ limit of the transition form factor agree in our calculations, which also provides a check that the multipole decomposition is implemented correctly.

\section{M1 strength in $^{48}$Ca}\label{sec:M1_48Ca}
\begin{figure*}[t!]
    \centering
    \includegraphics[width=\textwidth]{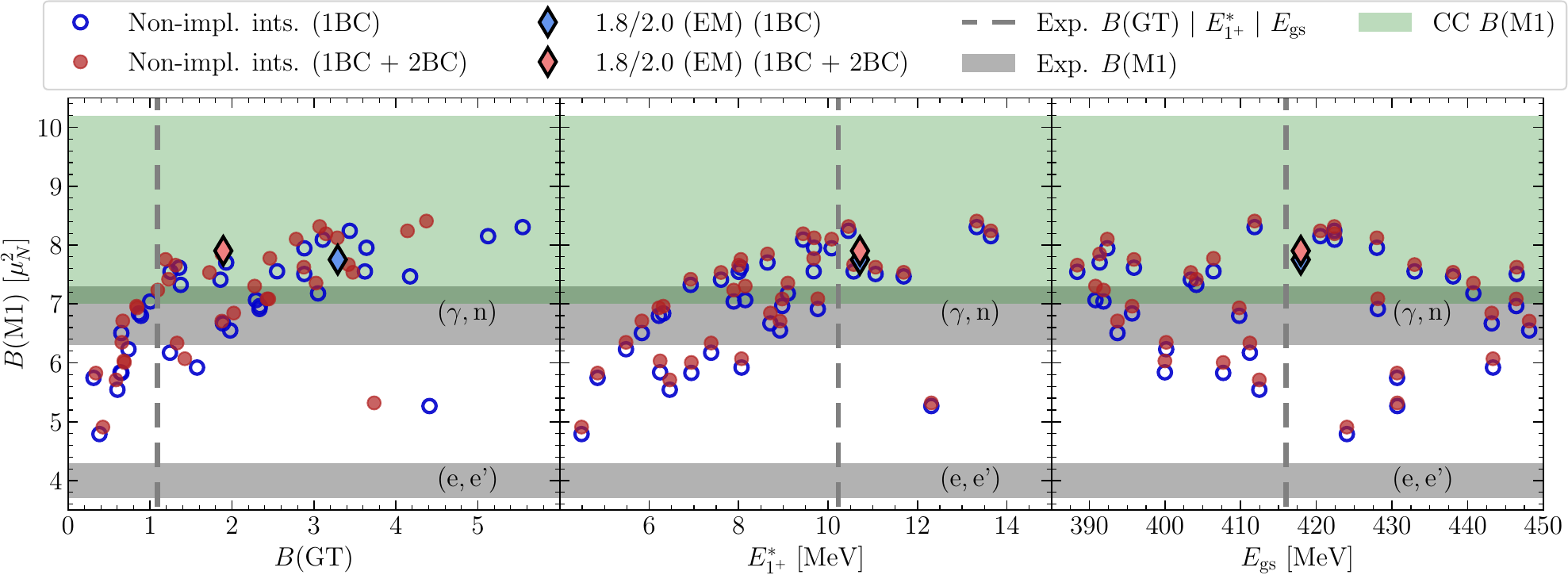}
    \caption{Correlations for $^{48}$Ca between the studied $B$(M1) and the largest $B$(GT) (see text for details) as well as between the $B$(M1) and the $1^+$ excitation energy $E^*_{1^+}$ and ground-state energy $E_\mathrm{gs}$ (in the left, middle and right panel, respectively). The gray bands show the $B(M1)$ to the 10.23\,MeV $1^{+}$ state measured with either the $(e,e')$ experiment \cite{Steffen1983} (lower gray band) or the $(\gamma,n)$ experiment \cite{Tompkins2011} (upper gray band). The dashed gray line in each panel shows the experimental $B$(GT) (from $^{48}$Ca($^3$He,t)$^{48}$Sc at 2.53\,MeV~\cite{Grewe2007}), $1^+$ excitation energy $E^*_{1^+}=10.23$\,MeV, and the ground-state energy $E_\mathrm{gs}$ of $^{48}$Ca~\cite{Nndc}. The green bar shows the prediction of the $B$(M1) from coupled-cluster calculations using different NN+3N interactions~\cite{Acharya2024}. Our results using the non-implausible interactions with 1BCs and 2BCs (or 1BCs only) are given by the red filled circles (blue open circles). The red and blue filled diamonds are the results using the 1.8/2.0 (EM) interaction with and without 2BCs, respectively.}
    \label{fig:B(M1)_corr}
\end{figure*}

Our results for the $B$(M1) in $^{48}$Ca are presented in Fig.~\ref{fig:B(M1)_corr} in comparison with experiment~\cite{Steffen1980, Tompkins2011} and coupled-cluster calculations using four different Hamiltonians~\cite{Acharya2024}. The three panels show the correlations of the $B$(M1) with the GT strength (left panel) with the $1^+$ excitation energy (middle panel), and with the ground-state energy of $^{48}$Ca (right panel). For the $B$(GT) we consider the transition with the largest strength (in experiment this is the $1^+$ state at 2.53\,MeV). The experimental $B$(GT) is taken from a $^{48}$Ca($^3$He,t)$^{48}$Sc charge-exchange reaction measurement~\cite{Grewe2007}, and the $1^+$ excitation energy and the ground-state energy from Ref.~\cite{Nndc}. Our VS-IMSRG results using the non-implausible interactions with 1BCs and 2BCs (or 1BCs only) are given by the red filled circles (blue open circles). The range of the results is expected to reflect a non-implausible range of the theoretical prediction. A full uncertainty quantification requires evaluations of uncertainties from other sources, such as chiral EFT and the many-body method, and is left for future work. In addition, we present our results using the 1.8/2.0 (EM) interaction with and without 2BCs (red and blue filled diamonds).

\subsection{Discussion of observables}

In all three panels of Fig.~\ref{fig:B(M1)_corr}, the ranges covered by the non-implausible interactions (with and without 2BCs) enclose the experimental $B$(GT), $1^+$ excitation energy, and ground-state energy. The $B$(M1) obtained from the $(\gamma,n)$ measurement is also covered by the results obtained with the non-implausible interactions. However, the $B$(M1) measured with the $(e,e')$ experiment lies outside of the theory range. Moreover, for the $B$(M1) strength of this transition, we find only a small effect from 2BCs, which increase the $B$(M1) (see also Fig.~\ref{fig:strength_ratios}). Our results for the 1.8/2.0~(EM) interaction lie within the range of the non-implausible interactions. Again, there is only a small increase from 2BCs for the $B$(M1). As expected for this interaction, the experimental ground-state energy and the $1^+$ excitation energy are very well reproduced. Our prediction favors larger $B$(M1) compared to the $(e,e')$ result and is in agreement with that from coupled-cluster calculations. While the 2BC contributions for the $B$(GT) are generally larger than for the $B$(M1), and generally decrease the GT strength, the effects for the non-implausible interactions are not as large as for the 1.8/2.0~(EM) interaction, where the 2BC contribution moves the $B$(GT) significantly closer towards experiment.

In the left and middle panel of Fig.~\ref{fig:B(M1)_corr}, there is a slight correlation between the observables, which can be quantified with the correlation coefficient $r$. The correlation between the $B$(M1) and $B$(GT) from the non-implausible interactions without 2BCs exhibits a correlation coefficient $r_\mathrm{1BC:M1,\,GT}=0.61$. The observed correlation is somewhat expected as both M1 and GT operators are dominated by the spin-isospin $\vec{\sigma} \vec{\tau}$ term. The same value is obtained for the correlation between the $B$(M1) and the excitation energy of the corresponding $1^+$ state ($r_\mathrm{1BC:M1,\,\mathit{E}_{1^+}}=0.61$).  
The 2BC contributions do not significantly change the correlations. This can be expected for the correlations with the excitation energy, since 2BCs have a very small impact on the $B$(M1) and none on the energy. Surprisingly, the correlation coefficient also does not change for the correlation between M1 and GT strengths ($r_\mathrm{1BC+2BC: M1,\,GT}=0.62$), even though 2BCs have a larger impact on the $B$(GT) strengths.

\subsection{Discussion of 2BCs}

\begin{figure}[t]
    \centering
    \includegraphics[width=0.8\linewidth]{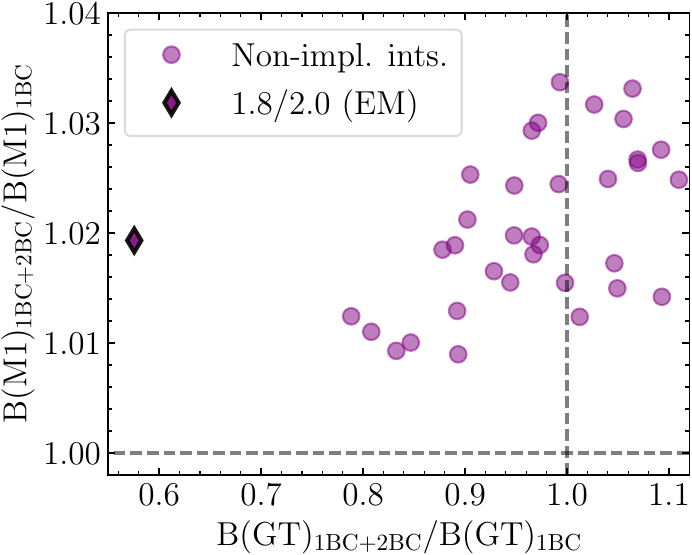}
    \caption{Ratio between the $B$(M1) and $B$(GT) strengths computed with and without 2BCs. The result with only 1BCs is denoted as $B$(M1/GT)$_{\rm 1BC}$. Likewise, the result with the 2BCs is given by $B$(M1/GT)$_{\rm 1BC+2BC}$. The circles show the results obtained with the non-implausible interactions, and the diamond is for the 1.8/2.0 (EM) interaction result.}
    \label{fig:strength_ratios}
\end{figure}

Next, we study  in more detail the impact of the vector and axial-vector 2BCs on the M1 and GT transitions. The ratio between the results computed with and without 2BCs, indicated by $B$(M1/GT)$_{\rm 1BC+2BC}$ and $B$(M1/GT)$_{\rm 1BC}$ is shown in Fig.~\ref{fig:strength_ratios}. The two dashed lines indicate that there is no contribution from 2BCs. For all results, 2BCs enhance the $B$(M1). The enhancement of the $B$(M1) is small, and does not exceed $4\%$. This is consistent with the effects of the 2BCs on the transition form factor at $Q \rightarrow 0$ illustrated in Fig.~\ref{fig:TFF}. 

Based on the calculations with the 1.8/2.0~(EM) interaction, we find that the small 2BCs effect for this M1 transition originates from a strong cancellation between the seagull and pion-in-flight contributions, corresponding to the first and second terms in the square bracket in Eq.~\eqref{eq:2BC}, respectively. Note that we did not observe such strong cancellation in the magnetic moment calculations of Ref.~\cite{Miyagi2024}, suggesting that the cancellation is accidental here. For the $^{48}$Ca($0^{+} \to 1^{+}$; 10.23\,MeV) transition, this may be an indication that one needs to include higher-order 1BC and 2BC contributions~\cite{Krebs2019,Krebs2020}, which include additional low-energy constants that would need to be fitted to electromagnetic observables. This is also needed to fully estimate the theoretical uncertainties.

In contrast, Fig.~\ref{fig:strength_ratios} shows that the inclusion of axial-vector 2BCs change significantly the 1BC results. The $B$(GT) using the 1.8/2.0~(EM) interaction is reduced by more than 40$\%$, whereas the $B$(M1) is only enhanced by less than $2\%$. For the non-implausible interactions, the reduction in $B$(GT) due to 2BCs is much less pronounced (smaller than $25\%$), and for some cases the GT strength even increases when 2BCs are included (up to $\sim 11\%$). We find that the results showing strong reduction from 2BCs tend to give large GT strength with 1BCs only, and that the 2BCs shift the results toward the experimental value. Likewise, the two smallest $B$(GT) from the set of non-implausible interactions are slightly enhanced by the 2BCs. Overall, this complex behavior of the axial-vector 2BCs results in a better description of the experimental $B$(GT), starting from the non-implausible samples.

As shown in Fig.~\ref{fig:strength_ratios}, 2BCs overall decrease $B$(GT) and slightly enhance $B$(M1). This is consistent with the study of $\beta$ decay, where it was shown that 2BCs and many-body correlations can explain the quenching puzzle in these transitions~\cite{Gysbers2019}. 
We have also investigated the role of 2BCs in the M1 transitions between the ground and other computed $1^+$ states. For most of the 34 non-implausible interactions, the inclusion of 2BCs enhances half or more calculated M1 strengths. Only for six interactions, the 2BCs enhance the $B$(M1) values in less than half of the transitions. For the 1.8/2.0~(EM) interaction, $80\%$ of the calculated states are enhanced by the 2BCs. And in all cases the absolute enhancement is small. This illustrates the different behavior of vector and axial-vector 2BCs for M1 and GT transitions, and confirms that our finding of small leading 2BC contributions to M1 transitions seems to be robust.

\section{M1 strength in $^{48}$Ti}\label{sec:M1_48Ti}
\begin{figure*}
    \centering
    \includegraphics[width=0.75\textwidth]{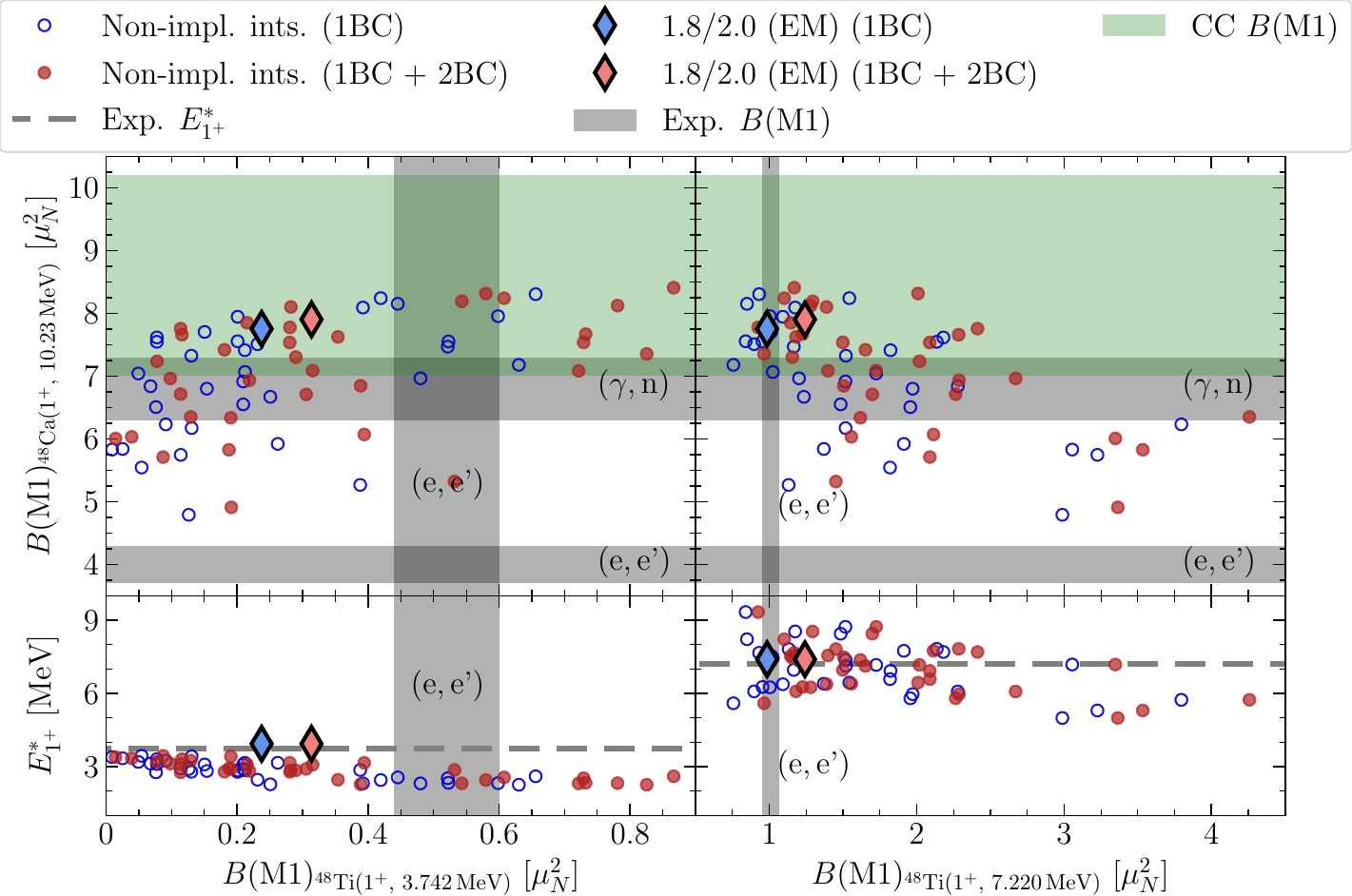}
    \caption{Correlation between the M1 strength $B$(M1) in $^{48}$Ca and $^{48}$Ti (top panels) and between the $1^+$ excitation energy $E^*_{1^+}$ and the M1 strength in $^{48}$Ti (bottom panels). The left (right) panels show the correlations with the M1 strength in $^{48}$Ti at 3.742\,MeV (7.22\,MeV) \cite{Guhr1990}. The gray bands show the M1 strength in $^{48}$Ca at 10.23\,MeV either measured with the $(e,e')$ experiment \cite{Steffen1983} (lower gray band) or the $(\gamma,n)$ experiment \cite{Tompkins2011} (upper gray band). The dashed grey lines and the vertical grey bands show the experimental values for the $1^+$ excitation energies and the $B$(M1) strengths in $^{48}$Ti \cite{Nndc,Guhr1990}. The green bar shows the prediction of the M1 strength from coupled-cluster calculations using different NN+3N interactions~\cite{Acharya2024}. Our results for the non-implausible interactions including only 1BCs (including 1BCs and 2BCs) are given by the blue open circles (red filled circles) and for the 1.8/2.0 (EM) interaction by the blue (red) filled diamonds.}
    \label{fig:B(M1)_corr_48Ti}
\end{figure*}

To validate our results for the M1 strength in $^{48}$Ca we additionally study M1 transitions in the neighboring nucleus $^{48}$Ti, which have been measured using inelastic electron scattering~\cite{Guhr1990}. Compared to $^{48}$Ca, the measured $B(\text{M1})$ strengths in $^{48}$Ti are more evenly distributed among $1^+$ states, and the identification between a calculated state and an experimental one becomes less clear. In order to simplify the assignment, we focus on the first excited $1^+$ state at 3.742\,MeV, which has the second strongest $B$(M1), and the $1^+$ state at 7.22\,MeV excitation energy, where the $B(\text{M1})$ is the strongest among all measured values.

We calculate the 20 lowest-energy $1^+$ states in $^{48}$Ti. To identify a calculated state with the experimental ones presented above, we use the following criteria. First, we tentatively assign the lowest $1^+$ state in our calculation to the  3.742\,MeV state. Second, as an additional indication, we explore the nucleon occupation numbers. In Ref.~\cite{Richter1990}, the 3.742\,MeV state is associated with a M1 form factor uniquely dominated by the $f_{7/2}$ single-particle orbital. Consistently with this, in our calculations the sum of proton and neutron occupation numbers in the $f_{7/2}$ orbital is usually the highest for the lowest-lying state, among all calculated $1^+$ states. This result holds for all but one non-implausible interaction, where the state with highest $f_{7/2}$ occupancy is the second-lowest $1^+$ state instead. Based on this, we identify the calculated state with the highest sum of $f_{7/2}$ occupation numbers with the experimental state at 3.742\,MeV. For the $1^+$ state at 7.22\,MeV, we follow the same criterion we employed for $^{48}$Ca, and identify this experimental state with the calculated one exhibiting the strongest $B(\text{M1})$.

Figure~\ref{fig:B(M1)_corr_48Ti} compares the results of our theoretical calculations for the energies and $B(\text{M1})$ strengths of the 3.742\,MeV and 7.22\,MeV excited states in $^{48}$Ti with the experimental ones from Refs.~\cite{Guhr1990,Nndc}. This shows that the measured excitation energies and M1 strengths for both states lie within the uncertainty range spanned by the theoretical predictions obtained with the non-implausible interactions. The 1.8/2.0~(EM) interaction underestimates the $B$(M1) of the state at 3.742\,MeV, but describes the stronger $B(\text{M1})$ of the state at 7.22\,MeV rather well.

Figure~\ref{fig:B(M1)_corr_48Ti} also compares the $B(\text{M1})$ strengths obtained with and without 2BCs for each nuclear Hamiltonian we used, similar to our study of the 10.23\,MeV state in $^{48}$Ca, which is also shown in this figure. While the effect of 2BCs also enhances the two $B(\text{M1})$ values studied in $^{48}$Ti, this effect is larger than in $^{48}$Ca: between 30\% and 65\% for the transition to the 3.742\,MeV, and between 6\% and 32\% for the 7.22\,MeV $B(\text{M1})$.

Finally, we have also explored the impact of 2BCs on the $B(\text{M1})$ strength for the 20 lowest-energy $1^+$ states in $^{48}$Ti. For 24 of the 34 non-implausible interactions the inclusion of 2BCs enhances the $B(\text{M1})$ in more than $80\%$ of the states. Only for one of the non-implausible interactions 2BCs increase the $B(\text{M1})$ for just $50\%$ of the calculated $1^+$ states. Likewise, for the 1.8/2.0~(EM) interaction, 2BCs enhance the M1 strength for $85\%$ of the calculated $1^+$ states.

\section{Summary and conclusions}\label{sec:conclusion}
We have derived and implemented a multipole decomposition of leading-order 2BCs to include them in {\it ab initio} calculations of medium-mass and heavy nuclei involving finite momentum transfers. As first applications, we have investigated the effects of 2BCs on the strong M1 transition at 10.23\,MeV in $^{48}$Ca, the related transition form factor, as well as the two strongest M1 transitions in $^{48}$Ti. To explore the Hamiltonian uncertainty, we have studied a broad set of non-implausible interactions as well as the 1.8/2.0 (EM) interaction, which are all based on chiral EFT.

For the transition form factor in $^{48}$Ca, we find overall a similar trend compared to experiment, but at low momentum transfer $Q \lesssim 50$\,MeV our results overestimate inelastic electron scattering data~\cite{Steffen1983}. The inclusion of 2BCs slightly enhances the discrepancy even more. For higher momentum transfer, there is overlap between experiment and theory, but the uncertainties from the experiment and the nuclear interactions considered are both large.

Our results for the $B$(M1) strength at 10.23\,MeV in $^{48}$Ca are also larger than the one extracted from the $(e,e')$ experiment~\cite{Steffen1980} but in agreement with the value obtained from the $(\gamma,n)$ experiment~\cite{Tompkins2011}. For the M1 strength, the inclusion of 2BCs slightly increases the strength, consistent with the findings of recent coupled-cluster calculations~\cite{Acharya2024}. However, for this transition in $^{48}$Ca, there are larger cancellations between the leading pion-in-flight and seagull 2BCs, so that future calculations including higher-order 2BCs are important. 

For validation, we have investigated two M1 transitions in $^{48}$Ti. For this nucleus, both $1^+$ excitation energies and M1 strengths~\cite{Guhr1990} are well described by our VS-IMSRG calculations within the range of Hamiltonians studied.

We have also studied axial-vector 2BCs for GT transitions in $^{48}$Ca. For GT transitions, the axial-vector 2BCs provide a larger contribution compared to the one of vector 2BC contributions to M1 transitions. Therefore, using similar quenching factors for M1 and GT transitions in phenomenological calculations is not supported from first principles.

Finally, the application of the multipole-decomposed 2BCs in calculations at finite momentum transfer opens a broad range of applications to very interesting electroweak processes and experiments involving medium-mass and heavy nuclei. This includes neutrinoless $\beta\beta$ decay, muon capture, as well as the scattering of weakly interacting massive particles and neutrinos off nuclei. 

\section{Data availability}

The data that support the findings of this article are openly available, including the python scripts for the figures \cite{Zenodo}.

\acknowledgments

We thank Michio Kohno, Peter von Neumann-Cosel, Thomas Papenbrock, and Atsushi Tamii for helpful discussions. This work was supported in part by the European Research Council (ERC) under the European Union’s Horizon 2020 research and innovation programme (Grant Agreement No.~101020842), by the JST ERATO Grant No.~JPMJER2304, Japan, by MCIN/AEI/10.13039/501100011033 from the grants: PID2023-147112NB-C22; CNS2022-135716 funded by the ``European Union NextGenerationEU/PRTR''; and CEX2019-000918-M to the ``Unit of Excellence Mar\'ia de Maeztu 2020-2023'' award to the Institute of Cosmos Sciences. This research in part used computational resources provided by Multidisciplinary Cooperative Research Program in Center for Computational Sciences, University of Tsukuba.

\appendix
\section{Multipole decomposition of leading two-body currents}\label{sec:multipoles}
The goal of this section is to obtain the expressions of the matrix element of the multipole components from the leading order two-body electromagnetic current.
We use the same notation as in Refs.~\cite{Krebs2019,Krebs2020}, and the coordinate-space current operator $\vec{j}_{\rm c}(\vec{x})$ is defined as
\begin{equation}
\vec{j}_{\rm c}(\vec{x}) = \int \frac{d\vec{Q}}{(2\pi)^{3}} e^{-i\vec{Q}\cdot \vec{x}} \vec{j}(\vec{Q}) \,.
\end{equation}
We have checked that the momentum-space multipole decomposition is equivalent to the well-known coordinate-space multipole decomposition found in Ref.~\cite{Walecka:1995mi}, through analyses with the leading 1BC operator.
To achieve a numerical calculation with a spherically formulated method, the current needs to be decomposed with vector spherical harmonics as
\begin{widetext}
\begin{equation}
\vec{j}(\vec{Q}) = 4\pi \sum_{\lambda\mu} i^{\lambda+1} \left[
L_{\lambda\mu}(Q)\vec{Y}^{*}_{\lambda\mu}(\uvec{Q}) 
+ T^{\rm el}_{\lambda\mu}(Q)\vec{\Psi}^{*}_{\lambda\mu}(\uvec{Q}) 
+ T^{\rm mag}_{\lambda\mu}(Q)\left[i\vec{\Phi}^{*}_{\lambda\mu}(\uvec{Q})\right]^{*}
\right],
\end{equation}
with the vector spherical harmonics
\begin{equation}
\vec{Y}_{\lambda\mu}(\uvec{Q}) = \uvec{Q} Y_{\lambda\mu}(\uvec{Q}), \quad
\vec{\Psi}_{\lambda\mu}(\uvec{Q}) = \sqrt{\frac{1}{\lambda(\lambda+1)}} Q \nabla_{\vec{Q}} Y_{\lambda\mu}(\uvec{Q}), \quad
\vec{\Phi}_{\lambda\mu}(\uvec{Q}) = \sqrt{\frac{1}{\lambda(\lambda+1)}} (\vec{Q} \times \nabla_{\vec{Q}}) Y_{\lambda\mu}(\uvec{Q}) 
\end{equation}
The vector spherical harmonics are orthonormal to each other:
\begin{equation}
\begin{aligned}
\int d\uvec{Q} \, \vec{Y}^{*}_{\lambda\mu}(\uvec{Q}) \cdot \vec{Y}_{\lambda'\mu'}(\uvec{Q}) &= 
\int d\uvec{Q} \, \vec{\Psi}^{*}_{\lambda\mu}(\uvec{Q}) \cdot \vec{\Psi}_{\lambda'\mu'}(\uvec{Q}) = 
\int d\uvec{Q} \, \vec{\Phi}^{*}_{\lambda\mu}(\uvec{Q}) \cdot \vec{\Phi}_{\lambda'\mu'}(\uvec{Q}) = \delta_{\lambda\lambda'} \delta_{\mu\mu'} \,, \\
\int d\uvec{Q} \, \vec{Y}^{*}_{\lambda\mu}(\uvec{Q}) \cdot \vec{\Psi}_{\lambda'\mu'}(\uvec{Q}) &=
\int d\uvec{Q} \, \vec{Y}^{*}_{\lambda\mu}(\uvec{Q}) \cdot \vec{\Phi}_{\lambda'\mu'}(\uvec{Q}) = 
\int d\uvec{Q} \, \vec{\Psi}^{*}_{\lambda\mu}(\uvec{Q}) \cdot \vec{\Phi}_{\lambda'\mu'}(\uvec{Q}) = 0 \,.
\end{aligned}
\end{equation}
The phases are chosen such that
\begin{equation}
\vec{Y}^{*}_{\lambda\mu}(\uvec{Q}) = (-1)^{\mu}\vec{Y}_{\lambda, -\mu}(\uvec{Q})\,, \quad
\vec{\Psi}^{*}_{\lambda\mu}(\uvec{Q}) = (-1)^{\mu}\vec{\Psi}_{\lambda, -\mu}(\uvec{Q})\,, \quad
\vec{\Phi}^{*}_{\lambda\mu}(\uvec{Q}) = (-1)^{\mu}\vec{\Phi}_{\lambda, -\mu}(\uvec{Q})\,, 
\end{equation}
as for the spherical harmonics.
Also, $\vec{Y}_{\lambda\mu}(\uvec{Q})$, $\vec{\Psi}_{\lambda\mu}(\uvec{Q})$, and $\vec{\Phi}_{\lambda\mu}(\uvec{Q})$ have $(-1)^{\lambda+1}$, $(-1)^{\lambda+1}$, and $(-1)^{\lambda}$ parities, respectively.
With the orthonormal condition, the $L_{\lambda\mu}(Q)$, $T^{\rm el}_{\lambda\mu}(Q)$, and $T^{\rm mag}_{\lambda\mu}(Q)$ multipole components are obtained as
\begin{equation}
\begin{aligned}
\label{eq:def_multipoles}
L_{\lambda\mu}(Q) &= \frac{(-i)^{\lambda+1}}{4\pi} \int d\uvec{Q} \,\vec{Y}_{\lambda\mu}(\uvec{Q}) \cdot \vec{j}(\vec{Q})\,, \\
T^{\rm el}_{\lambda\mu}(Q) &= \frac{(-i)^{\lambda+1}}{4\pi} \int d\uvec{Q} \, \vec{\Psi}_{\lambda\mu}(\uvec{Q}) \cdot \vec{j}(\vec{Q})\,,\\
T^{\rm mag}_{\lambda\mu}(Q) &= \frac{(-i)^{\lambda}}{4\pi} \int d\uvec{Q} \, \vec{\Phi}_{\lambda\mu}(\uvec{Q}) \cdot \vec{j}(\vec{Q})\,. 
\end{aligned}
\end{equation}
Note that the operator definition differs by $i$ from the usually used multipole components~\cite{Walecka:1995mi}.
This factor is chosen such that $L^{\dag}_{\lambda\mu}(Q) = L_{\lambda,-\mu}(Q)$, $T^{{\rm el}\; \dag}_{\lambda\mu}(Q) = T^{\rm el}_{\lambda,-\mu}(Q)$, and 
$T^{{\rm mag}\; \dag}_{\lambda\mu}(Q) = T^{\rm mag}_{\lambda,-\mu}(Q)$, as for the usual spherical tensor operators, i.e., one finds $\langle J||X_{\lambda} || J' \rangle = (-1)^{J'-J} \langle J'||X_{\lambda}||J\rangle^{*}$, where $X_{\lambda}$ is either $L_{\lambda}(Q)$, $T^{\rm el}_{\lambda}(Q)$, or $T^{\rm mag}_{\lambda}(Q)$, and $J'$ and $J$ are angular momenta.

Since $L_{\lambda\mu}(Q)$, $T^{\rm el}_{\lambda\mu}(Q)$, and $T^{\rm mag}_{\lambda\mu}(Q)$ are spherical tensors, it is enough to obtain the reduced matrix elements $\langle P'p'\alpha'|| X_{\lambda} || Pp\alpha \rangle$.
The calculations are tedious but straightforward, and one can find
\begin{equation}
\begin{aligned}
\label{eq:me_L_2bc}
\langle P' p' \alpha' || L_{\lambda}(Q) || Pp\alpha \rangle &= \frac{i^{L'+l'-L-l-\lambda-1}eg_{A}^{2}}{16 \pi F^{2}_{\pi} }i(\vec{\tau}_{1} \times \vec{\tau}_{2})_{z}
\\  & \times \left\{
\sqrt{\frac{\lambda}{[\lambda]}} \left[\mathcal{A}^{\lambda-1,\lambda}_{\alpha'\alpha}(P',p',P,p,Q)+2\mathcal{B}^{\lambda-1,\lambda}_{\alpha'\alpha}(P',p',P,p,Q)\right] 
\right. \\ & \left. \hspace{2em}- 
\sqrt{\frac{\lambda+1}{[\lambda]}} 
\left[\mathcal{A}^{\lambda+1,\lambda}_{\alpha'\alpha}(P',p',P,p,Q) + 2\mathcal{B}^{\lambda+1,\lambda}_{\alpha'\alpha}(P',p',P,p,Q)\right]
\right\},
\end{aligned}
\end{equation}
\begin{equation}
\begin{aligned}
\label{eq:me_Tel_2bc}
\langle P' p' \alpha' || T^{\rm el}_{\lambda}(Q) || Pp\alpha \rangle &= \frac{i^{L'+l'-L-l-\lambda-1}eg_{A}^{2}}{16 \pi F^{2}_{\pi} } i(\vec{\tau}_{1} \times \vec{\tau}_{2})_{z}
\\ & \times \left\{
\sqrt{\frac{\lambda+1}{[\lambda]}} \left[\mathcal{A}^{\lambda-1,\lambda}_{\alpha'\alpha}(P',p',P,p,Q)+2\mathcal{B}^{\lambda-1,\lambda}_{\alpha'\alpha}(P',p',P,p,Q)\right] 
\right. \\ & \left. \hspace{2em} + 
\sqrt{\frac{\lambda}{[\lambda]}} 
\left[\mathcal{A}^{\lambda+1,\lambda}_{\alpha'\alpha}(P',p',P,p,Q) + 2\mathcal{B}^{\lambda+1,\lambda}_{\alpha'\alpha}(P',p',P,p,Q)\right]
\right\},
\end{aligned}
\end{equation}
\begin{equation}
\begin{aligned}
\langle P' p' \alpha' || T^{\rm mag}_{\lambda}(Q) || Pp\alpha \rangle = -\frac{i^{L'+l'-L-l-\lambda-1}eg_{A}^{2}}{16 \pi F^{2}_{\pi} } i(\vec{\tau}_{1} \times \vec{\tau}_{2})_{z}
& \left[\mathcal{A}^{\lambda,\lambda}_{\alpha'\alpha}(P',p',P,p,Q)+2\mathcal{B}^{\lambda,\lambda}_{\alpha'\alpha}(P',p',P,p,Q)\right].
\end{aligned}
\end{equation}
Here, $[x]=2x+1$ is used.
The $\mathcal{A}^{\kappa,\lambda}_{\alpha'\alpha}(P',p',P,p,Q)$ and $\mathcal{B}^{\kappa,\lambda}_{\alpha'\alpha}(P',p',P,p,Q)$ functions are given by
\begin{equation}
\label{eq:functionA}
\begin{aligned}
\mathcal{A}^{\kappa,\lambda}_{\alpha'\alpha}(P',p',P,p,Q) &= (-1)^{\lambda+\kappa} 
\sqrt{4\pi[J'][J][\lambda][\kappa]}
\sum_{k_{1}+k_{2}=1} (-1)^{k_{2}} 
\left(\frac{1}{2}\right)^{k_{1}} \sqrt{[k_{1}]} 
\\ & \times \sum_{KN} 
\sqrt{[K][N]} 
\ninej{L'}{S'}{J'}{L}{S}{J}{N}{K}{\lambda}
\sixj{\kappa}{1}{\lambda}{K}{N}{1} 
 \langle S' || [\sigma_{1}\sigma_{2}]_{K} || S \rangle
 \\ & \times
 \sum_{X}(-1)^{X} 
 \sixj{\kappa}{k_{1}}{X}{k_{2}}{N}{1}
 \cg{\kappa}{k_{1}}{X}{0}{0}{0}
 \\ & \times
\mathcal{O}^{(k_{2}X)N}_{\alpha'\alpha}\left(P',p',P,p,Q,  
\frac{Q^{k_{1}} q^{k_{2}}}{q^{2}_{1}+m^{2}_{\pi}} - (-1)^{k_{2}+N} \frac{Q^{k_{1}} q^{k_{2}}}{q^{2}_{2}+m^{2}_{\pi}}\right),
\end{aligned}
\end{equation}
and 
\begin{equation}
\label{eq:functionB}
\begin{aligned}
    \mathcal{B}^{\kappa,\lambda}_{\alpha'\alpha}&(P',p',P,p,Q) = (-1)^{\lambda} 3 \sqrt{4\pi [J'][J][\lambda][\kappa]} \sum_{KN} (-1)^{K} \sqrt{[K][N]}
    \ninej{L'_{\rm tot}}{S'}{J'}{L_{\rm tot}}{S}{J}{N}{K}{\lambda} \langle S' || [\sigma_{1}\sigma_{2}]_{K} || S \rangle
    \\ & \times
    \sum_{k_{1}+k_{2}=1} \sum_{k_{3}+k_{4}=1} \left(\frac{1}{2} \right)^{k_{1}+k_{3}} (-1)^{k_{4}}
    \sum_{k_{13}k_{24}} \sqrt{[k_{13}][k_{24}]} \ninej{k_{1}}{k_{2}}{1}{k_{3}}{k_{4}}{1}{k_{13}}{k_{24}}{K} \cgz{k_{1}}{k_{3}}{k_{13}} \cgz{k_{2}}{k_{4}}{k_{24}}
    \\ & \times
    \sum_{X_{1}X_{2}} (-1)^{X_{1}+X_{2}}
    \ninej{\kappa}{1}{\lambda}{k_{13}}{k_{24}}{K}{X_{1}}{X_{2}}{N}
    \cgz{\kappa}{k_{13}}{X_{1}}
    \cgz{1}{k_{24}}{X_{2}} \mathcal{O}^{(X_{2}X_{1})N}_{\alpha'\alpha}\left(P',p',P,p,Q,  
\frac{Q^{k_{1}+k_{3}} q^{k_{2}+k_{4}+1}}
{(q^{2}_{1}+m^{2}_{\pi})(q^{2}_{2}+m^{2}_{\pi})}\right),
\end{aligned}
\end{equation}
using the Clebsch-Gordan coefficient $\cg{j_{1}}{j_{2}}{j}{m_{1}}{m_{2}}{m_{1}+m_{2}}$ and 6$j$- and 9$j$-symbols with the usual notation.
The orbital function $\mathcal{O}^{(\lambda_{1}\lambda_{2})\lambda}_{\alpha'\alpha}[P',p',P,p,Q,f(Q, q, \cos\theta)]$ is defined as
\begin{equation}
\begin{aligned}
\mathcal{O}^{(\lambda_{1}\lambda_{2})\lambda}_{\alpha'\alpha}&[P',p',P,p,Q,f(Q, q, \cos\theta)] = 
(2\pi)^{3} \frac{(-1)^{L'+l'}}{P'p'PpQ} 
\sqrt{[\lambda_{1}][\lambda_{2}][\lambda][L'][L]}
  \\ & \times\sum_{\bar{L}\bar{l}\kappa} (-1)^{\kappa} \sqrt{[\bar{L}][\bar{l}]}
 \cgz{\bar{l}}{\lambda_{1}}{\kappa}
\cgz{\bar{L}}{\lambda_{2}}{\kappa} 
\sixj{\bar{l}}{\lambda_{1}}{\kappa}{\lambda_{2}}{\bar{L}}{\lambda}
\ninej{L'_{\rm cm}}{l'}{L'}{L_{\rm cm}}{l}{L}{\bar{L}}{\bar{l}}{\lambda}
\\ & \times
\left. \mathcal{Y}^{\bar{L}}_{L'L}(\widehat{\vec{P}+Q\vec{e}_{z}}, \uvec{P})
\right|_{\phi_{\vec{P}} = 0, \, \vec{P}\cdot \vec{e}_{z} = \frac{P'^{2}-P^{2}-Q^{2}}{2Q}}
\int^{p'+p}_{|p'-p|} dq q \left. \mathcal{Y}^{\bar{l}}_{l'l}(\widehat{\vec{p}+q\vec{e}_{z}}, \uvec{p})\right|_{\phi_{\vec{p}} = 0, \, \vec{p}\cdot \vec{e}_{z} = \frac{p'^{2}-p^{2}-q^{2}}{2q}}
\\ & \times
\int d(\cos\theta) 
f(Q,q,\cos\theta) P_{\kappa}(\cos\theta) \,.
\end{aligned}
\end{equation}
\end{widetext}
Here, $\theta$ is the angle between $\vec{Q}$ and $\vec{q}$.
In the derivation, we exploited that the function $f(Q,q,\vec{Q}\cdot \vec{q})$ only depends on $Q$, $q$, and $\theta$.
Once $\langle P'p'\alpha'|| X_{\lambda} || Pp\alpha \rangle$ is computed, one can obtain the matrix element $\langle N'n' \alpha' || X_{\lambda} || Nn\alpha \rangle$ in the harmonic-oscillator basis through integral, Eq.~\eqref{eq:MomToHO}. 
As a final step, one needs to transform the matrix element to the coupled single-particle basis $|ab: J_{ab} \rangle$.
This can be done with Eq.~\eqref{eq:TMTrans}, and the overlap $\langle Nn\alpha | ab : J_{ab} \rangle$ can be written as
\begin{equation}
\label{eq:TMbracket}
\begin{aligned}
\langle Nn\alpha & |ab:J_{ab} \rangle 
 = f_{ab}\sqrt{[j_{a}][j_{b}][L_{\rm tot}][S]}
\\ & \times
\ninej{l_{a}}{1/2}{j_{a}}{l_{b}}{1/2}{j_{b}}{L_{\rm tot}}{S}{J}
\\ & \times
\langle NLnl:L_{\rm tot} | n_{a}l_{a}n_{b}l_{b}:L_{\rm tot} \rangle.
\end{aligned}
\end{equation}
Here, $\langle NLnl:L_{\rm tot} | n_{a}l_{a}n_{b}l_{b}:L_{\rm tot} \rangle$ is the harmonic-oscillator bracket with the notation in Ref.~\cite{Kamuntavicius2001}.
Also, $f_{ab}$ is defined in the neutron-proton (n-p) basis as
\begin{equation}
f_{ab} = \left\{
\begin{array}{cc}
\sqrt{\frac{1}{2(1+\tilde{\delta}_{ab})}} [1 + (-1)^{l+S}] \,, &\text{pp or nn} \\ 
1 \,, & \text{pn}
\end{array}
\right.
\end{equation}
with $\tilde{\delta}_{ab} = \delta_{n_{a}n_{b}} \delta_{l_{a}l_{b}} \delta_{j_{a}j_{b}}$.
We note that the reduced matrix elements of $L_{\lambda\mu}(Q)$ and $T^{\rm el}_{\lambda\mu}(Q)$ are purely imaginary, and those of $T^{\rm mag}_{\lambda\mu}(Q)$ are always real. 
Therefore, the contributions from $L_{\lambda\mu}(Q)$ and $T^{\rm el}_{\lambda\mu}(Q)$ vanish for elastic scattering.
The actual implementation can be found in the \texttt{NuHamil} code~\cite{Miyagi2023}.

\bibliography{bib.bib}

\begin{thebibliography}{61}%
\makeatletter
\providecommand \@ifxundefined [1]{%
 \@ifx{#1\undefined}
}%
\providecommand \@ifnum [1]{%
 \ifnum #1\expandafter \@firstoftwo
 \else \expandafter \@secondoftwo
 \fi
}%
\providecommand \@ifx [1]{%
 \ifx #1\expandafter \@firstoftwo
 \else \expandafter \@secondoftwo
 \fi
}%
\providecommand \natexlab [1]{#1}%
\providecommand \enquote  [1]{``#1''}%
\providecommand \bibnamefont  [1]{#1}%
\providecommand \bibfnamefont [1]{#1}%
\providecommand \citenamefont [1]{#1}%
\providecommand \href@noop [0]{\@secondoftwo}%
\providecommand \href [0]{\begingroup \@sanitize@url \@href}%
\providecommand \@href[1]{\@@startlink{#1}\@@href}%
\providecommand \@@href[1]{\endgroup#1\@@endlink}%
\providecommand \@sanitize@url [0]{\catcode `\\12\catcode `\$12\catcode `\&12\catcode `\#12\catcode `\^12\catcode `\_12\catcode `\%12\relax}%
\providecommand \@@startlink[1]{}%
\providecommand \@@endlink[0]{}%
\providecommand \url  [0]{\begingroup\@sanitize@url \@url }%
\providecommand \@url [1]{\endgroup\@href {#1}{\urlprefix }}%
\providecommand \urlprefix  [0]{URL }%
\providecommand \Eprint [0]{\href }%
\providecommand \doibase [0]{https://doi.org/}%
\providecommand \selectlanguage [0]{\@gobble}%
\providecommand \bibinfo  [0]{\@secondoftwo}%
\providecommand \bibfield  [0]{\@secondoftwo}%
\providecommand \translation [1]{[#1]}%
\providecommand \BibitemOpen [0]{}%
\providecommand \bibitemStop [0]{}%
\providecommand \bibitemNoStop [0]{.\EOS\space}%
\providecommand \EOS [0]{\spacefactor3000\relax}%
\providecommand \BibitemShut  [1]{\csname bibitem#1\endcsname}%
\let\auto@bib@innerbib\@empty
\bibitem [{\citenamefont {Nakamura}\ \emph {et~al.}(2001)\citenamefont {Nakamura}, \citenamefont {Sato}, \citenamefont {Gudkov},\ and\ \citenamefont {Kubodera}}]{Nakamura:2000vp}%
  \BibitemOpen
  \bibfield  {author} {\bibinfo {author} {\bibfnamefont {S.}~\bibnamefont {Nakamura}}, \bibinfo {author} {\bibfnamefont {T.}~\bibnamefont {Sato}}, \bibinfo {author} {\bibfnamefont {V.~P.}\ \bibnamefont {Gudkov}},\ and\ \bibinfo {author} {\bibfnamefont {K.}~\bibnamefont {Kubodera}},\ }\bibfield  {title} {\bibinfo {title} {{Neutrino reactions on deuteron}},\ }\href {https://doi.org/doi.org/10.1103/PhysRevC.63.034617} {\bibfield  {journal} {\bibinfo  {journal} {Phys. Rev. C}\ }\textbf {\bibinfo {volume} {63}},\ \bibinfo {pages} {034617} (\bibinfo {year} {2001})},\ \bibinfo {note} {[\href{https://doi.org/10.1103/PhysRevC.73.049904}{Erratum: Phys. Rev. C 73, 049904 (2006)}]}\BibitemShut {NoStop}%
\bibitem [{\citenamefont {Butler}\ \emph {et~al.}(2001)\citenamefont {Butler}, \citenamefont {Chen},\ and\ \citenamefont {Kong}}]{Butler2001}%
  \BibitemOpen
  \bibfield  {author} {\bibinfo {author} {\bibfnamefont {M.}~\bibnamefont {Butler}}, \bibinfo {author} {\bibfnamefont {J.-W.}\ \bibnamefont {Chen}},\ and\ \bibinfo {author} {\bibfnamefont {X.}~\bibnamefont {Kong}},\ }\bibfield  {title} {\bibinfo {title} {Neutrino-deuteron scattering in effective field theory at next-to-next-to-leading order},\ }\href {https://doi.org/10.1103/PhysRevC.63.035501} {\bibfield  {journal} {\bibinfo  {journal} {Phys. Rev. C}\ }\textbf {\bibinfo {volume} {63}},\ \bibinfo {pages} {035501} (\bibinfo {year} {2001})}\BibitemShut {NoStop}%
\bibitem [{\citenamefont {Gazit}\ \emph {et~al.}(2009)\citenamefont {Gazit}, \citenamefont {Quaglioni},\ and\ \citenamefont {Navr{\'a}til}}]{Gazit2009}%
  \BibitemOpen
  \bibfield  {author} {\bibinfo {author} {\bibfnamefont {D.}~\bibnamefont {Gazit}}, \bibinfo {author} {\bibfnamefont {S.}~\bibnamefont {Quaglioni}},\ and\ \bibinfo {author} {\bibfnamefont {P.}~\bibnamefont {Navr{\'a}til}},\ }\bibfield  {title} {\bibinfo {title} {Three-{{Nucleon Low-Energy Constants}} from the {{Consistency}} of {{Interactions}} and {{Currents}} in {{Chiral Effective Field Theory}}},\ }\href {https://doi.org/10.1103/PhysRevLett.103.102502} {\bibfield  {journal} {\bibinfo  {journal} {Phys. Rev. Lett.}\ }\textbf {\bibinfo {volume} {103}},\ \bibinfo {pages} {102502} (\bibinfo {year} {2009})}\BibitemShut {NoStop}%
\bibitem [{\citenamefont {Bacca}\ and\ \citenamefont {Pastore}(2014)}]{Bacca:2014tla}%
  \BibitemOpen
  \bibfield  {author} {\bibinfo {author} {\bibfnamefont {S.}~\bibnamefont {Bacca}}\ and\ \bibinfo {author} {\bibfnamefont {S.}~\bibnamefont {Pastore}},\ }\bibfield  {title} {\bibinfo {title} {{Electromagnetic reactions on light nuclei}},\ }\href {https://doi.org/10.1088/0954-3899/41/12/123002} {\bibfield  {journal} {\bibinfo  {journal} {J. Phys. G}\ }\textbf {\bibinfo {volume} {41}},\ \bibinfo {pages} {123002} (\bibinfo {year} {2014})}\BibitemShut {NoStop}%
\bibitem [{\citenamefont {Marcucci}\ \emph {et~al.}(2016)\citenamefont {Marcucci}, \citenamefont {Gross}, \citenamefont {Pena}, \citenamefont {Piarulli}, \citenamefont {Schiavilla}, \citenamefont {Sick}, \citenamefont {Stadler}, \citenamefont {Van~Orden},\ and\ \citenamefont {Viviani}}]{Marcucci:2015rca}%
  \BibitemOpen
  \bibfield  {author} {\bibinfo {author} {\bibfnamefont {L.~E.}\ \bibnamefont {Marcucci}}, \bibinfo {author} {\bibfnamefont {F.}~\bibnamefont {Gross}}, \bibinfo {author} {\bibfnamefont {M.~T.}\ \bibnamefont {Pena}}, \bibinfo {author} {\bibfnamefont {M.}~\bibnamefont {Piarulli}}, \bibinfo {author} {\bibfnamefont {R.}~\bibnamefont {Schiavilla}}, \bibinfo {author} {\bibfnamefont {I.}~\bibnamefont {Sick}}, \bibinfo {author} {\bibfnamefont {A.}~\bibnamefont {Stadler}}, \bibinfo {author} {\bibfnamefont {J.~W.}\ \bibnamefont {Van~Orden}},\ and\ \bibinfo {author} {\bibfnamefont {M.}~\bibnamefont {Viviani}},\ }\bibfield  {title} {\bibinfo {title} {{Electromagnetic Structure of Few-Nucleon Ground States}},\ }\href {https://doi.org/10.1088/0954-3899/43/2/023002} {\bibfield  {journal} {\bibinfo  {journal} {J. Phys. G}\ }\textbf {\bibinfo {volume} {43}},\ \bibinfo {pages} {023002} (\bibinfo {year} {2016})}\BibitemShut {NoStop}%
\bibitem [{\citenamefont {Lynn}\ \emph {et~al.}(2019)\citenamefont {Lynn}, \citenamefont {Tews}, \citenamefont {Gandolfi},\ and\ \citenamefont {Lovato}}]{Lynn:2019rdt}%
  \BibitemOpen
  \bibfield  {author} {\bibinfo {author} {\bibfnamefont {J.~E.}\ \bibnamefont {Lynn}}, \bibinfo {author} {\bibfnamefont {I.}~\bibnamefont {Tews}}, \bibinfo {author} {\bibfnamefont {S.}~\bibnamefont {Gandolfi}},\ and\ \bibinfo {author} {\bibfnamefont {A.}~\bibnamefont {Lovato}},\ }\bibfield  {title} {\bibinfo {title} {{Quantum Monte Carlo Methods in Nuclear Physics: Recent Advances}},\ }\href {https://doi.org/10.1146/annurev-nucl-101918-023600} {\bibfield  {journal} {\bibinfo  {journal} {Ann. Rev. Nucl. Part. Sci.}\ }\textbf {\bibinfo {volume} {69}},\ \bibinfo {pages} {279} (\bibinfo {year} {2019})}\BibitemShut {NoStop}%
\bibitem [{\citenamefont {Gnech}\ and\ \citenamefont {Schiavilla}(2022)}]{Gnech:2022vwr}%
  \BibitemOpen
  \bibfield  {author} {\bibinfo {author} {\bibfnamefont {A.}~\bibnamefont {Gnech}}\ and\ \bibinfo {author} {\bibfnamefont {R.}~\bibnamefont {Schiavilla}},\ }\bibfield  {title} {\bibinfo {title} {{Magnetic structure of few-nucleon systems at high momentum transfers in a chiral effective field theory approach}},\ }\href {https://doi.org/10.1103/PhysRevC.106.044001} {\bibfield  {journal} {\bibinfo  {journal} {Phys. Rev. C}\ }\textbf {\bibinfo {volume} {106}},\ \bibinfo {pages} {044001} (\bibinfo {year} {2022})}\BibitemShut {NoStop}%
\bibitem [{\citenamefont {Seutin}\ \emph {et~al.}(2023)\citenamefont {Seutin}, \citenamefont {Hernandez}, \citenamefont {Miyagi}, \citenamefont {Bacca}, \citenamefont {Hebeler}, \citenamefont {K\"onig},\ and\ \citenamefont {Schwenk}}]{Seutin:2023grs}%
  \BibitemOpen
  \bibfield  {author} {\bibinfo {author} {\bibfnamefont {R.}~\bibnamefont {Seutin}}, \bibinfo {author} {\bibfnamefont {O.~J.}\ \bibnamefont {Hernandez}}, \bibinfo {author} {\bibfnamefont {T.}~\bibnamefont {Miyagi}}, \bibinfo {author} {\bibfnamefont {S.}~\bibnamefont {Bacca}}, \bibinfo {author} {\bibfnamefont {K.}~\bibnamefont {Hebeler}}, \bibinfo {author} {\bibfnamefont {S.}~\bibnamefont {K\"onig}},\ and\ \bibinfo {author} {\bibfnamefont {A.}~\bibnamefont {Schwenk}},\ }\bibfield  {title} {\bibinfo {title} {{Magnetic dipole operator from chiral effective field theory for many-body expansion methods}},\ }\href {https://doi.org/10.1103/PhysRevC.108.054005} {\bibfield  {journal} {\bibinfo  {journal} {Phys. Rev. C}\ }\textbf {\bibinfo {volume} {108}},\ \bibinfo {pages} {054005} (\bibinfo {year} {2023})}\BibitemShut {NoStop}%
\bibitem [{\citenamefont {Martin}\ \emph {et~al.}(2023)\citenamefont {Martin}, \citenamefont {Novario}, \citenamefont {Lonardoni}, \citenamefont {Carlson}, \citenamefont {Gandolfi},\ and\ \citenamefont {Tews}}]{Martin:2023dhl}%
  \BibitemOpen
  \bibfield  {author} {\bibinfo {author} {\bibfnamefont {J.~D.}\ \bibnamefont {Martin}}, \bibinfo {author} {\bibfnamefont {S.~J.}\ \bibnamefont {Novario}}, \bibinfo {author} {\bibfnamefont {D.}~\bibnamefont {Lonardoni}}, \bibinfo {author} {\bibfnamefont {J.}~\bibnamefont {Carlson}}, \bibinfo {author} {\bibfnamefont {S.}~\bibnamefont {Gandolfi}},\ and\ \bibinfo {author} {\bibfnamefont {I.}~\bibnamefont {Tews}},\ }\bibfield  {title} {\bibinfo {title} {{Auxiliary field diffusion Monte Carlo calculations of magnetic moments of light nuclei with chiral effective field theory interactions}},\ }\href {https://doi.org/10.1103/PhysRevC.108.L031304} {\bibfield  {journal} {\bibinfo  {journal} {Phys. Rev. C}\ }\textbf {\bibinfo {volume} {108}},\ \bibinfo {pages} {L031304} (\bibinfo {year} {2023})}\BibitemShut {NoStop}%
\bibitem [{\citenamefont {Pal}\ \emph {et~al.}(2023)\citenamefont {Pal}, \citenamefont {Sarker}, \citenamefont {Fasano}, \citenamefont {Maris}, \citenamefont {Vary}, \citenamefont {Caprio},\ and\ \citenamefont {Basili}}]{Pal:2023gll}%
  \BibitemOpen
  \bibfield  {author} {\bibinfo {author} {\bibfnamefont {S.}~\bibnamefont {Pal}}, \bibinfo {author} {\bibfnamefont {S.}~\bibnamefont {Sarker}}, \bibinfo {author} {\bibfnamefont {P.~J.}\ \bibnamefont {Fasano}}, \bibinfo {author} {\bibfnamefont {P.}~\bibnamefont {Maris}}, \bibinfo {author} {\bibfnamefont {J.~P.}\ \bibnamefont {Vary}}, \bibinfo {author} {\bibfnamefont {M.~A.}\ \bibnamefont {Caprio}},\ and\ \bibinfo {author} {\bibfnamefont {R.~A.~M.}\ \bibnamefont {Basili}},\ }\bibfield  {title} {\bibinfo {title} {{Magnetic moments of $A=3$ nuclei obtained from chiral effective field theory operators}},\ }\href {https://doi.org/10.1103/PhysRevC.108.024001} {\bibfield  {journal} {\bibinfo  {journal} {Phys. Rev. C}\ }\textbf {\bibinfo {volume} {108}},\ \bibinfo {pages} {024001} (\bibinfo {year} {2023})}\BibitemShut {NoStop}%
\bibitem [{\citenamefont {{Chambers-Wall}}\ \emph {et~al.}(2024)\citenamefont {{Chambers-Wall}}, \citenamefont {Gnech}, \citenamefont {King}, \citenamefont {Pastore}, \citenamefont {Piarulli}, \citenamefont {Schiavilla},\ and\ \citenamefont {Wiringa}}]{Chambers-Wall2024}%
  \BibitemOpen
  \bibfield  {author} {\bibinfo {author} {\bibfnamefont {G.}~\bibnamefont {{Chambers-Wall}}}, \bibinfo {author} {\bibfnamefont {A.}~\bibnamefont {Gnech}}, \bibinfo {author} {\bibfnamefont {G.~B.}\ \bibnamefont {King}}, \bibinfo {author} {\bibfnamefont {S.}~\bibnamefont {Pastore}}, \bibinfo {author} {\bibfnamefont {M.}~\bibnamefont {Piarulli}}, \bibinfo {author} {\bibfnamefont {R.}~\bibnamefont {Schiavilla}},\ and\ \bibinfo {author} {\bibfnamefont {R.~B.}\ \bibnamefont {Wiringa}},\ }\bibfield  {title} {\bibinfo {title} {Quantum {{Monte Carlo Calculations}} of {{Magnetic Form Factors}} in {{Light Nuclei}}},\ }\href {https://doi.org/10.1103/PhysRevLett.133.212501} {\bibfield  {journal} {\bibinfo  {journal} {Phys. Rev. Lett.}\ }\textbf {\bibinfo {volume} {133}},\ \bibinfo {pages} {212501} (\bibinfo {year} {2024})}\BibitemShut {NoStop}%
\bibitem [{\citenamefont {Gysbers}\ \emph {et~al.}(2019)\citenamefont {Gysbers}, \citenamefont {Hagen}, \citenamefont {Holt}, \citenamefont {Jansen}, \citenamefont {Morris}, \citenamefont {Navr{\'a}til}, \citenamefont {Papenbrock}, \citenamefont {Quaglioni}, \citenamefont {Schwenk}, \citenamefont {Stroberg},\ and\ \citenamefont {Wendt}}]{Gysbers2019}%
  \BibitemOpen
  \bibfield  {author} {\bibinfo {author} {\bibfnamefont {P.}~\bibnamefont {Gysbers}}, \bibinfo {author} {\bibfnamefont {G.}~\bibnamefont {Hagen}}, \bibinfo {author} {\bibfnamefont {J.~D.}\ \bibnamefont {Holt}}, \bibinfo {author} {\bibfnamefont {G.~R.}\ \bibnamefont {Jansen}}, \bibinfo {author} {\bibfnamefont {T.~D.}\ \bibnamefont {Morris}}, \bibinfo {author} {\bibfnamefont {P.}~\bibnamefont {Navr{\'a}til}}, \bibinfo {author} {\bibfnamefont {T.}~\bibnamefont {Papenbrock}}, \bibinfo {author} {\bibfnamefont {S.}~\bibnamefont {Quaglioni}}, \bibinfo {author} {\bibfnamefont {A.}~\bibnamefont {Schwenk}}, \bibinfo {author} {\bibfnamefont {S.~R.}\ \bibnamefont {Stroberg}},\ and\ \bibinfo {author} {\bibfnamefont {K.~A.}\ \bibnamefont {Wendt}},\ }\bibfield  {title} {\bibinfo {title} {Discrepancy between experimental and theoretical {$\beta$}-decay rates resolved from first principles},\ }\href {https://doi.org/10.1038/s41567-019-0450-7} {\bibfield  {journal} {\bibinfo  {journal} {Nature Phys.}\ }\textbf {\bibinfo
  {volume} {15}},\ \bibinfo {pages} {428} (\bibinfo {year} {2019})}\BibitemShut {NoStop}%
\bibitem [{\citenamefont {King}\ \emph {et~al.}(2020)\citenamefont {King}, \citenamefont {Andreoli}, \citenamefont {Pastore}, \citenamefont {Piarulli}, \citenamefont {Schiavilla}, \citenamefont {Wiringa}, \citenamefont {Carlson},\ and\ \citenamefont {Gandolfi}}]{King2020}%
  \BibitemOpen
  \bibfield  {author} {\bibinfo {author} {\bibfnamefont {G.~B.}\ \bibnamefont {King}}, \bibinfo {author} {\bibfnamefont {L.}~\bibnamefont {Andreoli}}, \bibinfo {author} {\bibfnamefont {S.}~\bibnamefont {Pastore}}, \bibinfo {author} {\bibfnamefont {M.}~\bibnamefont {Piarulli}}, \bibinfo {author} {\bibfnamefont {R.}~\bibnamefont {Schiavilla}}, \bibinfo {author} {\bibfnamefont {R.~B.}\ \bibnamefont {Wiringa}}, \bibinfo {author} {\bibfnamefont {J.}~\bibnamefont {Carlson}},\ and\ \bibinfo {author} {\bibfnamefont {S.}~\bibnamefont {Gandolfi}},\ }\bibfield  {title} {\bibinfo {title} {Chiral effective field theory calculations of weak transitions in light nuclei},\ }\href {https://doi.org/10.1103/PhysRevC.102.025501} {\bibfield  {journal} {\bibinfo  {journal} {Phys. Rev. C}\ }\textbf {\bibinfo {volume} {102}},\ \bibinfo {pages} {025501} (\bibinfo {year} {2020})}\BibitemShut {NoStop}%
\bibitem [{\citenamefont {King}\ \emph {et~al.}(2023)\citenamefont {King}, \citenamefont {Baroni}, \citenamefont {Cirigliano}, \citenamefont {Gandolfi}, \citenamefont {Hayen}, \citenamefont {Mereghetti}, \citenamefont {Pastore},\ and\ \citenamefont {Piarulli}}]{King:2022zkz}%
  \BibitemOpen
  \bibfield  {author} {\bibinfo {author} {\bibfnamefont {G.~B.}\ \bibnamefont {King}}, \bibinfo {author} {\bibfnamefont {A.}~\bibnamefont {Baroni}}, \bibinfo {author} {\bibfnamefont {V.}~\bibnamefont {Cirigliano}}, \bibinfo {author} {\bibfnamefont {S.}~\bibnamefont {Gandolfi}}, \bibinfo {author} {\bibfnamefont {L.}~\bibnamefont {Hayen}}, \bibinfo {author} {\bibfnamefont {E.}~\bibnamefont {Mereghetti}}, \bibinfo {author} {\bibfnamefont {S.}~\bibnamefont {Pastore}},\ and\ \bibinfo {author} {\bibfnamefont {M.}~\bibnamefont {Piarulli}},\ }\bibfield  {title} {\bibinfo {title} {{Ab initio calculation of the \ensuremath{\beta}-decay spectrum of \ensuremath{^6}He}},\ }\href {https://doi.org/10.1103/PhysRevC.107.015503} {\bibfield  {journal} {\bibinfo  {journal} {Phys. Rev. C}\ }\textbf {\bibinfo {volume} {107}},\ \bibinfo {pages} {015503} (\bibinfo {year} {2023})}\BibitemShut {NoStop}%
\bibitem [{\citenamefont {King}\ \emph {et~al.}(2022)\citenamefont {King}, \citenamefont {Pastore}, \citenamefont {Piarulli},\ and\ \citenamefont {Schiavilla}}]{King2022}%
  \BibitemOpen
  \bibfield  {author} {\bibinfo {author} {\bibfnamefont {G.~B.}\ \bibnamefont {King}}, \bibinfo {author} {\bibfnamefont {S.}~\bibnamefont {Pastore}}, \bibinfo {author} {\bibfnamefont {M.}~\bibnamefont {Piarulli}},\ and\ \bibinfo {author} {\bibfnamefont {R.}~\bibnamefont {Schiavilla}},\ }\bibfield  {title} {\bibinfo {title} {{Partial Muon Capture Rates in $A = 3$ and $A = 6$ Nuclei with Chiral Effective Field Theory}},\ }\href {https://doi.org/10.1103/PhysRevC.105.L042501} {\bibfield  {journal} {\bibinfo  {journal} {Phys. Rev. C}\ }\textbf {\bibinfo {volume} {105}},\ \bibinfo {pages} {L042501} (\bibinfo {year} {2022})}\BibitemShut {NoStop}%
\bibitem [{\citenamefont {Andreoli}\ \emph {et~al.}(2024)\citenamefont {Andreoli}, \citenamefont {King}, \citenamefont {Pastore}, \citenamefont {Piarulli}, \citenamefont {Carlson}, \citenamefont {Gandolfi},\ and\ \citenamefont {Wiringa}}]{Andreoli:2024ovl}%
  \BibitemOpen
  \bibfield  {author} {\bibinfo {author} {\bibfnamefont {L.}~\bibnamefont {Andreoli}}, \bibinfo {author} {\bibfnamefont {G.~B.}\ \bibnamefont {King}}, \bibinfo {author} {\bibfnamefont {S.}~\bibnamefont {Pastore}}, \bibinfo {author} {\bibfnamefont {M.}~\bibnamefont {Piarulli}}, \bibinfo {author} {\bibfnamefont {J.}~\bibnamefont {Carlson}}, \bibinfo {author} {\bibfnamefont {S.}~\bibnamefont {Gandolfi}},\ and\ \bibinfo {author} {\bibfnamefont {R.~B.}\ \bibnamefont {Wiringa}},\ }\bibfield  {title} {\bibinfo {title} {{Quantum Monte Carlo calculations of electron scattering from \ensuremath{^{12}}C in the short-time approximation}},\ }\href {https://doi.org/10.1103/PhysRevC.110.064004} {\bibfield  {journal} {\bibinfo  {journal} {Phys. Rev. C}\ }\textbf {\bibinfo {volume} {110}},\ \bibinfo {pages} {064004} (\bibinfo {year} {2024})}\BibitemShut {NoStop}%
\bibitem [{\citenamefont {Lovato}\ \emph {et~al.}(2020)\citenamefont {Lovato}, \citenamefont {Carlson}, \citenamefont {Gandolfi}, \citenamefont {Rocco},\ and\ \citenamefont {Schiavilla}}]{Lovato:2020kba}%
  \BibitemOpen
  \bibfield  {author} {\bibinfo {author} {\bibfnamefont {A.}~\bibnamefont {Lovato}}, \bibinfo {author} {\bibfnamefont {J.}~\bibnamefont {Carlson}}, \bibinfo {author} {\bibfnamefont {S.}~\bibnamefont {Gandolfi}}, \bibinfo {author} {\bibfnamefont {N.}~\bibnamefont {Rocco}},\ and\ \bibinfo {author} {\bibfnamefont {R.}~\bibnamefont {Schiavilla}},\ }\bibfield  {title} {\bibinfo {title} {{Ab initio study of $\boldsymbol{(\nu_\ell,\ell^-)}$ and $\boldsymbol{(\overline{\nu}_\ell,\ell^+)}$ inclusive scattering in $^{12}$C: confronting the MiniBooNE and T2K CCQE data}},\ }\href {https://doi.org/10.1103/PhysRevX.10.031068} {\bibfield  {journal} {\bibinfo  {journal} {Phys. Rev. X}\ }\textbf {\bibinfo {volume} {10}},\ \bibinfo {pages} {031068} (\bibinfo {year} {2020})}\BibitemShut {NoStop}%
\bibitem [{\citenamefont {Miyagi}\ \emph {et~al.}(2024)\citenamefont {Miyagi}, \citenamefont {Cao}, \citenamefont {Seutin}, \citenamefont {Bacca}, \citenamefont {Ruiz}, \citenamefont {Hebeler}, \citenamefont {Holt},\ and\ \citenamefont {Schwenk}}]{Miyagi2024}%
  \BibitemOpen
  \bibfield  {author} {\bibinfo {author} {\bibfnamefont {T.}~\bibnamefont {Miyagi}}, \bibinfo {author} {\bibfnamefont {X.}~\bibnamefont {Cao}}, \bibinfo {author} {\bibfnamefont {R.}~\bibnamefont {Seutin}}, \bibinfo {author} {\bibfnamefont {S.}~\bibnamefont {Bacca}}, \bibinfo {author} {\bibfnamefont {R.~F.~G.}\ \bibnamefont {Ruiz}}, \bibinfo {author} {\bibfnamefont {K.}~\bibnamefont {Hebeler}}, \bibinfo {author} {\bibfnamefont {J.~D.}\ \bibnamefont {Holt}},\ and\ \bibinfo {author} {\bibfnamefont {A.}~\bibnamefont {Schwenk}},\ }\bibfield  {title} {\bibinfo {title} {Impact of {{Two-Body Currents}} on {{Magnetic Dipole Moments}} of {{Nuclei}}},\ }\href {https://doi.org/10.1103/PhysRevLett.132.232503} {\bibfield  {journal} {\bibinfo  {journal} {Phys. Rev. Lett.}\ }\textbf {\bibinfo {volume} {132}},\ \bibinfo {pages} {232503} (\bibinfo {year} {2024})}\BibitemShut {NoStop}%
\bibitem [{\citenamefont {Acharya}\ \emph {et~al.}(2024)\citenamefont {Acharya}, \citenamefont {Hu}, \citenamefont {Bacca}, \citenamefont {Hagen}, \citenamefont {Navr{\'a}til},\ and\ \citenamefont {Papenbrock}}]{Acharya2024}%
  \BibitemOpen
  \bibfield  {author} {\bibinfo {author} {\bibfnamefont {B.}~\bibnamefont {Acharya}}, \bibinfo {author} {\bibfnamefont {B.~S.}\ \bibnamefont {Hu}}, \bibinfo {author} {\bibfnamefont {S.}~\bibnamefont {Bacca}}, \bibinfo {author} {\bibfnamefont {G.}~\bibnamefont {Hagen}}, \bibinfo {author} {\bibfnamefont {P.}~\bibnamefont {Navr{\'a}til}},\ and\ \bibinfo {author} {\bibfnamefont {T.}~\bibnamefont {Papenbrock}},\ }\bibfield  {title} {\bibinfo {title} {Magnetic {{Dipole Transition}} in \ensuremath{^{48}}{{Ca}}},\ }\href {https://doi.org/10.1103/PhysRevLett.132.232504} {\bibfield  {journal} {\bibinfo  {journal} {Phys. Rev. Lett.}\ }\textbf {\bibinfo {volume} {132}},\ \bibinfo {pages} {232504} (\bibinfo {year} {2024})}\BibitemShut {NoStop}%
\bibitem [{\citenamefont {Jokiniemi}\ \emph {et~al.}(2023{\natexlab{a}})\citenamefont {Jokiniemi}, \citenamefont {Miyagi}, \citenamefont {Stroberg}, \citenamefont {Holt}, \citenamefont {Kotila},\ and\ \citenamefont {Suhonen}}]{Jokiniemi:2021qqg}%
  \BibitemOpen
  \bibfield  {author} {\bibinfo {author} {\bibfnamefont {L.}~\bibnamefont {Jokiniemi}}, \bibinfo {author} {\bibfnamefont {T.}~\bibnamefont {Miyagi}}, \bibinfo {author} {\bibfnamefont {S.~R.}\ \bibnamefont {Stroberg}}, \bibinfo {author} {\bibfnamefont {J.~D.}\ \bibnamefont {Holt}}, \bibinfo {author} {\bibfnamefont {J.}~\bibnamefont {Kotila}},\ and\ \bibinfo {author} {\bibfnamefont {J.}~\bibnamefont {Suhonen}},\ }\bibfield  {title} {\bibinfo {title} {{Ab initio calculation of muon capture on $^{24}$Mg}},\ }\href {https://doi.org/10.1103/PhysRevC.107.014327} {\bibfield  {journal} {\bibinfo  {journal} {Phys. Rev. C}\ }\textbf {\bibinfo {volume} {107}},\ \bibinfo {pages} {014327} (\bibinfo {year} {2023}{\natexlab{a}})}\BibitemShut {NoStop}%
\bibitem [{\citenamefont {Gimeno}\ \emph {et~al.}(2023)\citenamefont {Gimeno}, \citenamefont {Jokiniemi}, \citenamefont {Kotila}, \citenamefont {Ramalho},\ and\ \citenamefont {Suhonen}}]{Gimeno2023}%
  \BibitemOpen
  \bibfield  {author} {\bibinfo {author} {\bibfnamefont {P.}~\bibnamefont {Gimeno}}, \bibinfo {author} {\bibfnamefont {L.}~\bibnamefont {Jokiniemi}}, \bibinfo {author} {\bibfnamefont {J.}~\bibnamefont {Kotila}}, \bibinfo {author} {\bibfnamefont {M.}~\bibnamefont {Ramalho}},\ and\ \bibinfo {author} {\bibfnamefont {J.}~\bibnamefont {Suhonen}},\ }\bibfield  {title} {\bibinfo {title} {Ordinary {{Muon Capture}} on \ensuremath{^{136}}{{Ba}}: {{Comparative Study Using}} the {{Shell Model}} and {{pnQRPA}}},\ }\href {https://doi.org/10.3390/universe9060270} {\bibfield  {journal} {\bibinfo  {journal} {Universe}\ }\textbf {\bibinfo {volume} {9}},\ \bibinfo {pages} {270} (\bibinfo {year} {2023})}\BibitemShut {NoStop}%
\bibitem [{\citenamefont {Jokiniemi}\ \emph {et~al.}(2024)\citenamefont {Jokiniemi}, \citenamefont {Navr{\'a}til}, \citenamefont {Kotila},\ and\ \citenamefont {Kravvaris}}]{Jokiniemi2024}%
  \BibitemOpen
  \bibfield  {author} {\bibinfo {author} {\bibfnamefont {L.}~\bibnamefont {Jokiniemi}}, \bibinfo {author} {\bibfnamefont {P.}~\bibnamefont {Navr{\'a}til}}, \bibinfo {author} {\bibfnamefont {J.}~\bibnamefont {Kotila}},\ and\ \bibinfo {author} {\bibfnamefont {K.}~\bibnamefont {Kravvaris}},\ }\bibfield  {title} {\bibinfo {title} {Muon capture on \ensuremath{^6}{{Li}}, \ensuremath{^{12}}{{C}}, and \ensuremath{^{16}}{{O}} from {\emph{ab initio}} nuclear theory},\ }\href {https://doi.org/10.1103/PhysRevC.109.065501} {\bibfield  {journal} {\bibinfo  {journal} {Phys. Rev. C}\ }\textbf {\bibinfo {volume} {109}},\ \bibinfo {pages} {065501} (\bibinfo {year} {2024})}\BibitemShut {NoStop}%
\bibitem [{\citenamefont {Hoferichter}\ \emph {et~al.}(2020)\citenamefont {Hoferichter}, \citenamefont {Men{\'e}ndez},\ and\ \citenamefont {Schwenk}}]{Hoferichter2020}%
  \BibitemOpen
  \bibfield  {author} {\bibinfo {author} {\bibfnamefont {M.}~\bibnamefont {Hoferichter}}, \bibinfo {author} {\bibfnamefont {J.}~\bibnamefont {Men{\'e}ndez}},\ and\ \bibinfo {author} {\bibfnamefont {A.}~\bibnamefont {Schwenk}},\ }\bibfield  {title} {\bibinfo {title} {Coherent elastic neutrino-nucleus scattering: {{EFT}} analysis and nuclear responses},\ }\href {https://doi.org/10.1103/PhysRevD.102.074018} {\bibfield  {journal} {\bibinfo  {journal} {Phys. Rev. D}\ }\textbf {\bibinfo {volume} {102}},\ \bibinfo {pages} {074018} (\bibinfo {year} {2020})}\BibitemShut {NoStop}%
\bibitem [{\citenamefont {Men{\'e}ndez}\ \emph {et~al.}(2011)\citenamefont {Men{\'e}ndez}, \citenamefont {Gazit},\ and\ \citenamefont {Schwenk}}]{Menendez2011}%
  \BibitemOpen
  \bibfield  {author} {\bibinfo {author} {\bibfnamefont {J.}~\bibnamefont {Men{\'e}ndez}}, \bibinfo {author} {\bibfnamefont {D.}~\bibnamefont {Gazit}},\ and\ \bibinfo {author} {\bibfnamefont {A.}~\bibnamefont {Schwenk}},\ }\bibfield  {title} {\bibinfo {title} {Chiral {{Two-Body Currents}} in {{Nuclei}}: {{Gamow-Teller Transitions}} and {{Neutrinoless Double-Beta Decay}}},\ }\href {https://doi.org/10.1103/PhysRevLett.107.062501} {\bibfield  {journal} {\bibinfo  {journal} {Phys. Rev. Lett.}\ }\textbf {\bibinfo {volume} {107}},\ \bibinfo {pages} {062501} (\bibinfo {year} {2011})}\BibitemShut {NoStop}%
\bibitem [{\citenamefont {Engel}\ \emph {et~al.}(2014)\citenamefont {Engel}, \citenamefont {{\v S}imkovic},\ and\ \citenamefont {Vogel}}]{Engel2014}%
  \BibitemOpen
  \bibfield  {author} {\bibinfo {author} {\bibfnamefont {J.}~\bibnamefont {Engel}}, \bibinfo {author} {\bibfnamefont {F.}~\bibnamefont {{\v S}imkovic}},\ and\ \bibinfo {author} {\bibfnamefont {P.}~\bibnamefont {Vogel}},\ }\bibfield  {title} {\bibinfo {title} {Chiral two-body currents and neutrinoless double-{$\beta$} decay in the quasiparticle random-phase approximation},\ }\href {https://doi.org/10.1103/PhysRevC.89.064308} {\bibfield  {journal} {\bibinfo  {journal} {Phys. Rev. C}\ }\textbf {\bibinfo {volume} {89}},\ \bibinfo {pages} {064308} (\bibinfo {year} {2014})}\BibitemShut {NoStop}%
\bibitem [{\citenamefont {Jokiniemi}\ \emph {et~al.}(2023{\natexlab{b}})\citenamefont {Jokiniemi}, \citenamefont {Romeo}, \citenamefont {Soriano},\ and\ \citenamefont {Men\'endez}}]{Jokiniemi:2022ayc}%
  \BibitemOpen
  \bibfield  {author} {\bibinfo {author} {\bibfnamefont {L.}~\bibnamefont {Jokiniemi}}, \bibinfo {author} {\bibfnamefont {B.}~\bibnamefont {Romeo}}, \bibinfo {author} {\bibfnamefont {P.}~\bibnamefont {Soriano}},\ and\ \bibinfo {author} {\bibfnamefont {J.}~\bibnamefont {Men\'endez}},\ }\bibfield  {title} {\bibinfo {title} {{Neutrinoless \ensuremath{\beta}\ensuremath{\beta}-decay nuclear matrix elements from two-neutrino \ensuremath{\beta}\ensuremath{\beta}-decay data}},\ }\href {https://doi.org/10.1103/PhysRevC.107.044305} {\bibfield  {journal} {\bibinfo  {journal} {Phys. Rev. C}\ }\textbf {\bibinfo {volume} {107}},\ \bibinfo {pages} {044305} (\bibinfo {year} {2023}{\natexlab{b}})}\BibitemShut {NoStop}%
\bibitem [{\citenamefont {Men{\'e}ndez}\ \emph {et~al.}(2012)\citenamefont {Men{\'e}ndez}, \citenamefont {Gazit},\ and\ \citenamefont {Schwenk}}]{Menendez2012a}%
  \BibitemOpen
  \bibfield  {author} {\bibinfo {author} {\bibfnamefont {J.}~\bibnamefont {Men{\'e}ndez}}, \bibinfo {author} {\bibfnamefont {D.}~\bibnamefont {Gazit}},\ and\ \bibinfo {author} {\bibfnamefont {A.}~\bibnamefont {Schwenk}},\ }\bibfield  {title} {\bibinfo {title} {Spin-dependent {{WIMP}} scattering off nuclei},\ }\href {https://doi.org/10.1103/PhysRevD.86.103511} {\bibfield  {journal} {\bibinfo  {journal} {Phys. Rev. D}\ }\textbf {\bibinfo {volume} {86}},\ \bibinfo {pages} {103511} (\bibinfo {year} {2012})}\BibitemShut {NoStop}%
\bibitem [{\citenamefont {Klos}\ \emph {et~al.}(2013)\citenamefont {Klos}, \citenamefont {Men{\'e}ndez}, \citenamefont {Gazit},\ and\ \citenamefont {Schwenk}}]{Klos2013a}%
  \BibitemOpen
  \bibfield  {author} {\bibinfo {author} {\bibfnamefont {P.}~\bibnamefont {Klos}}, \bibinfo {author} {\bibfnamefont {J.}~\bibnamefont {Men{\'e}ndez}}, \bibinfo {author} {\bibfnamefont {D.}~\bibnamefont {Gazit}},\ and\ \bibinfo {author} {\bibfnamefont {A.}~\bibnamefont {Schwenk}},\ }\bibfield  {title} {\bibinfo {title} {Large-scale nuclear structure calculations for spin-dependent {{WIMP}} scattering with chiral effective field theory currents},\ }\href {https://doi.org/10.1103/PhysRevD.88.083516} {\bibfield  {journal} {\bibinfo  {journal} {Phys. Rev. D}\ }\textbf {\bibinfo {volume} {88}},\ \bibinfo {pages} {083516} (\bibinfo {year} {2013})}\BibitemShut {NoStop}%
\bibitem [{\citenamefont {Baudis}\ \emph {et~al.}(2013)\citenamefont {Baudis}, \citenamefont {Kessler}, \citenamefont {Klos}, \citenamefont {Lang}, \citenamefont {Men{\'e}ndez}, \citenamefont {Reichard},\ and\ \citenamefont {Schwenk}}]{Baudis2013}%
  \BibitemOpen
  \bibfield  {author} {\bibinfo {author} {\bibfnamefont {L.}~\bibnamefont {Baudis}}, \bibinfo {author} {\bibfnamefont {G.}~\bibnamefont {Kessler}}, \bibinfo {author} {\bibfnamefont {P.}~\bibnamefont {Klos}}, \bibinfo {author} {\bibfnamefont {R.~F.}\ \bibnamefont {Lang}}, \bibinfo {author} {\bibfnamefont {J.}~\bibnamefont {Men{\'e}ndez}}, \bibinfo {author} {\bibfnamefont {S.}~\bibnamefont {Reichard}},\ and\ \bibinfo {author} {\bibfnamefont {A.}~\bibnamefont {Schwenk}},\ }\bibfield  {title} {\bibinfo {title} {Signatures of dark matter scattering inelastically off nuclei},\ }\href {https://doi.org/10.1103/PhysRevD.88.115014} {\bibfield  {journal} {\bibinfo  {journal} {Phys. Rev. D}\ }\textbf {\bibinfo {volume} {88}},\ \bibinfo {pages} {115014} (\bibinfo {year} {2013})}\BibitemShut {NoStop}%
\bibitem [{\citenamefont {Hoferichter}\ \emph {et~al.}(2016)\citenamefont {Hoferichter}, \citenamefont {Klos}, \citenamefont {Men\'endez},\ and\ \citenamefont {Schwenk}}]{Hoferichter:2016nvd}%
  \BibitemOpen
  \bibfield  {author} {\bibinfo {author} {\bibfnamefont {M.}~\bibnamefont {Hoferichter}}, \bibinfo {author} {\bibfnamefont {P.}~\bibnamefont {Klos}}, \bibinfo {author} {\bibfnamefont {J.}~\bibnamefont {Men\'endez}},\ and\ \bibinfo {author} {\bibfnamefont {A.}~\bibnamefont {Schwenk}},\ }\bibfield  {title} {\bibinfo {title} {{Analysis strategies for general spin-independent WIMP-nucleus scattering}},\ }\href {https://doi.org/10.1103/PhysRevD.94.063505} {\bibfield  {journal} {\bibinfo  {journal} {Phys. Rev. D}\ }\textbf {\bibinfo {volume} {94}},\ \bibinfo {pages} {063505} (\bibinfo {year} {2016})}\BibitemShut {NoStop}%
\bibitem [{\citenamefont {Hoferichter}\ \emph {et~al.}(2019)\citenamefont {Hoferichter}, \citenamefont {Klos}, \citenamefont {Men{\'e}ndez},\ and\ \citenamefont {Schwenk}}]{Hoferichter2019}%
  \BibitemOpen
  \bibfield  {author} {\bibinfo {author} {\bibfnamefont {M.}~\bibnamefont {Hoferichter}}, \bibinfo {author} {\bibfnamefont {P.}~\bibnamefont {Klos}}, \bibinfo {author} {\bibfnamefont {J.}~\bibnamefont {Men{\'e}ndez}},\ and\ \bibinfo {author} {\bibfnamefont {A.}~\bibnamefont {Schwenk}},\ }\bibfield  {title} {\bibinfo {title} {Nuclear structure factors for general spin-independent {{WIMP-nucleus}} scattering},\ }\href {https://doi.org/10.1103/PhysRevD.99.055031} {\bibfield  {journal} {\bibinfo  {journal} {Phys. Rev. D}\ }\textbf {\bibinfo {volume} {99}},\ \bibinfo {pages} {055031} (\bibinfo {year} {2019})}\BibitemShut {NoStop}%
\bibitem [{\citenamefont {Engel}\ and\ \citenamefont {Men{\'e}ndez}(2017)}]{Engel2017a}%
  \BibitemOpen
  \bibfield  {author} {\bibinfo {author} {\bibfnamefont {J.}~\bibnamefont {Engel}}\ and\ \bibinfo {author} {\bibfnamefont {J.}~\bibnamefont {Men{\'e}ndez}},\ }\bibfield  {title} {\bibinfo {title} {Status and future of nuclear matrix elements for neutrinoless double-beta decay: A review},\ }\href {https://doi.org/10.1088/1361-6633/aa5bc5} {\bibfield  {journal} {\bibinfo  {journal} {Rept. Prog. Phys.}\ }\textbf {\bibinfo {volume} {80}},\ \bibinfo {pages} {046301} (\bibinfo {year} {2017})}\BibitemShut {NoStop}%
\bibitem [{\citenamefont {Aalbers}\ \emph {et~al.}(2023)\citenamefont {Aalbers} \emph {et~al.}}]{Aalbers:2022dzr}%
  \BibitemOpen
  \bibfield  {author} {\bibinfo {author} {\bibfnamefont {J.}~\bibnamefont {Aalbers}} \emph {et~al.},\ }\bibfield  {title} {\bibinfo {title} {{A next-generation liquid xenon observatory for dark matter and neutrino physics}},\ }\href {https://doi.org/10.1088/1361-6471/ac841a} {\bibfield  {journal} {\bibinfo  {journal} {J. Phys. G}\ }\textbf {\bibinfo {volume} {50}},\ \bibinfo {pages} {013001} (\bibinfo {year} {2023})}\BibitemShut {NoStop}%
\bibitem [{\citenamefont {Stroberg}\ \emph {et~al.}(2017)\citenamefont {Stroberg}, \citenamefont {Calci}, \citenamefont {Hergert}, \citenamefont {Holt}, \citenamefont {Bogner}, \citenamefont {Roth},\ and\ \citenamefont {Schwenk}}]{Stroberg:2016ung}%
  \BibitemOpen
  \bibfield  {author} {\bibinfo {author} {\bibfnamefont {S.~R.}\ \bibnamefont {Stroberg}}, \bibinfo {author} {\bibfnamefont {A.}~\bibnamefont {Calci}}, \bibinfo {author} {\bibfnamefont {H.}~\bibnamefont {Hergert}}, \bibinfo {author} {\bibfnamefont {J.~D.}\ \bibnamefont {Holt}}, \bibinfo {author} {\bibfnamefont {S.~K.}\ \bibnamefont {Bogner}}, \bibinfo {author} {\bibfnamefont {R.}~\bibnamefont {Roth}},\ and\ \bibinfo {author} {\bibfnamefont {A.}~\bibnamefont {Schwenk}},\ }\bibfield  {title} {\bibinfo {title} {{A nucleus-dependent valence-space approach to nuclear structure}},\ }\href {https://doi.org/10.1103/PhysRevLett.118.032502} {\bibfield  {journal} {\bibinfo  {journal} {Phys. Rev. Lett.}\ }\textbf {\bibinfo {volume} {118}},\ \bibinfo {pages} {032502} (\bibinfo {year} {2017})}\BibitemShut {NoStop}%
\bibitem [{\citenamefont {Morris}\ \emph {et~al.}(2018)\citenamefont {Morris}, \citenamefont {Simonis}, \citenamefont {Stroberg}, \citenamefont {Stumpf}, \citenamefont {Hagen}, \citenamefont {Holt}, \citenamefont {Jansen}, \citenamefont {Papenbrock}, \citenamefont {Roth},\ and\ \citenamefont {Schwenk}}]{Morris:2017vxi}%
  \BibitemOpen
  \bibfield  {author} {\bibinfo {author} {\bibfnamefont {T.~D.}\ \bibnamefont {Morris}}, \bibinfo {author} {\bibfnamefont {J.}~\bibnamefont {Simonis}}, \bibinfo {author} {\bibfnamefont {S.~R.}\ \bibnamefont {Stroberg}}, \bibinfo {author} {\bibfnamefont {C.}~\bibnamefont {Stumpf}}, \bibinfo {author} {\bibfnamefont {G.}~\bibnamefont {Hagen}}, \bibinfo {author} {\bibfnamefont {J.~D.}\ \bibnamefont {Holt}}, \bibinfo {author} {\bibfnamefont {G.~R.}\ \bibnamefont {Jansen}}, \bibinfo {author} {\bibfnamefont {T.}~\bibnamefont {Papenbrock}}, \bibinfo {author} {\bibfnamefont {R.}~\bibnamefont {Roth}},\ and\ \bibinfo {author} {\bibfnamefont {A.}~\bibnamefont {Schwenk}},\ }\bibfield  {title} {\bibinfo {title} {{Structure of the lightest tin isotopes}},\ }\href {https://doi.org/10.1103/PhysRevLett.120.152503} {\bibfield  {journal} {\bibinfo  {journal} {Phys. Rev. Lett.}\ }\textbf {\bibinfo {volume} {120}},\ \bibinfo {pages} {152503} (\bibinfo {year} {2018})}\BibitemShut {NoStop}%
\bibitem [{\citenamefont {Stroberg}\ \emph {et~al.}(2019)\citenamefont {Stroberg}, \citenamefont {Bogner}, \citenamefont {Hergert},\ and\ \citenamefont {Holt}}]{Stroberg:2019mxo}%
  \BibitemOpen
  \bibfield  {author} {\bibinfo {author} {\bibfnamefont {S.~R.}\ \bibnamefont {Stroberg}}, \bibinfo {author} {\bibfnamefont {S.~K.}\ \bibnamefont {Bogner}}, \bibinfo {author} {\bibfnamefont {H.}~\bibnamefont {Hergert}},\ and\ \bibinfo {author} {\bibfnamefont {J.~D.}\ \bibnamefont {Holt}},\ }\bibfield  {title} {\bibinfo {title} {{Nonempirical Interactions for the Nuclear Shell Model: An Update}},\ }\href {https://doi.org/10.1146/annurev-nucl-101917-021120} {\bibfield  {journal} {\bibinfo  {journal} {Ann. Rev. Nucl. Part. Sci.}\ }\textbf {\bibinfo {volume} {69}},\ \bibinfo {pages} {307} (\bibinfo {year} {2019})}\BibitemShut {NoStop}%
\bibitem [{\citenamefont {Stroberg}\ \emph {et~al.}(2021)\citenamefont {Stroberg}, \citenamefont {Holt}, \citenamefont {Schwenk},\ and\ \citenamefont {Simonis}}]{Stroberg:2019bch}%
  \BibitemOpen
  \bibfield  {author} {\bibinfo {author} {\bibfnamefont {S.~R.}\ \bibnamefont {Stroberg}}, \bibinfo {author} {\bibfnamefont {J.~D.}\ \bibnamefont {Holt}}, \bibinfo {author} {\bibfnamefont {A.}~\bibnamefont {Schwenk}},\ and\ \bibinfo {author} {\bibfnamefont {J.}~\bibnamefont {Simonis}},\ }\bibfield  {title} {\bibinfo {title} {{Ab Initio Limits of Atomic Nuclei}},\ }\href {https://doi.org/10.1103/PhysRevLett.126.022501} {\bibfield  {journal} {\bibinfo  {journal} {Phys. Rev. Lett.}\ }\textbf {\bibinfo {volume} {126}},\ \bibinfo {pages} {022501} (\bibinfo {year} {2021})}\BibitemShut {NoStop}%
\bibitem [{\citenamefont {Miyagi}\ \emph {et~al.}(2022)\citenamefont {Miyagi}, \citenamefont {Stroberg}, \citenamefont {Navr\'atil}, \citenamefont {Hebeler},\ and\ \citenamefont {Holt}}]{Miyagi:2021pdc}%
  \BibitemOpen
  \bibfield  {author} {\bibinfo {author} {\bibfnamefont {T.}~\bibnamefont {Miyagi}}, \bibinfo {author} {\bibfnamefont {S.~R.}\ \bibnamefont {Stroberg}}, \bibinfo {author} {\bibfnamefont {P.}~\bibnamefont {Navr\'atil}}, \bibinfo {author} {\bibfnamefont {K.}~\bibnamefont {Hebeler}},\ and\ \bibinfo {author} {\bibfnamefont {J.~D.}\ \bibnamefont {Holt}},\ }\bibfield  {title} {\bibinfo {title} {{Converged ab initio calculations of heavy nuclei}},\ }\href {https://doi.org/10.1103/PhysRevC.105.014302} {\bibfield  {journal} {\bibinfo  {journal} {Phys. Rev. C}\ }\textbf {\bibinfo {volume} {105}},\ \bibinfo {pages} {014302} (\bibinfo {year} {2022})}\BibitemShut {NoStop}%
\bibitem [{\citenamefont {Hebeler}\ \emph {et~al.}(2023)\citenamefont {Hebeler}, \citenamefont {Durant}, \citenamefont {Hoppe}, \citenamefont {Heinz}, \citenamefont {Schwenk}, \citenamefont {Simonis},\ and\ \citenamefont {Tichai}}]{Hebeler:2022aui}%
  \BibitemOpen
  \bibfield  {author} {\bibinfo {author} {\bibfnamefont {K.}~\bibnamefont {Hebeler}}, \bibinfo {author} {\bibfnamefont {V.}~\bibnamefont {Durant}}, \bibinfo {author} {\bibfnamefont {J.}~\bibnamefont {Hoppe}}, \bibinfo {author} {\bibfnamefont {M.}~\bibnamefont {Heinz}}, \bibinfo {author} {\bibfnamefont {A.}~\bibnamefont {Schwenk}}, \bibinfo {author} {\bibfnamefont {J.}~\bibnamefont {Simonis}},\ and\ \bibinfo {author} {\bibfnamefont {A.}~\bibnamefont {Tichai}},\ }\bibfield  {title} {\bibinfo {title} {{Normal ordering of three-nucleon interactions for ab initio calculations of heavy nuclei}},\ }\href {https://doi.org/10.1103/PhysRevC.107.024310} {\bibfield  {journal} {\bibinfo  {journal} {Phys. Rev. C}\ }\textbf {\bibinfo {volume} {107}},\ \bibinfo {pages} {024310} (\bibinfo {year} {2023})}\BibitemShut {NoStop}%
\bibitem [{\citenamefont {Heinz}\ \emph {et~al.}(2025)\citenamefont {Heinz}, \citenamefont {Miyagi}, \citenamefont {Stroberg}, \citenamefont {Tichai}, \citenamefont {Hebeler},\ and\ \citenamefont {Schwenk}}]{Heinz:2024juw}%
  \BibitemOpen
  \bibfield  {author} {\bibinfo {author} {\bibfnamefont {M.}~\bibnamefont {Heinz}}, \bibinfo {author} {\bibfnamefont {T.}~\bibnamefont {Miyagi}}, \bibinfo {author} {\bibfnamefont {S.~R.}\ \bibnamefont {Stroberg}}, \bibinfo {author} {\bibfnamefont {A.}~\bibnamefont {Tichai}}, \bibinfo {author} {\bibfnamefont {K.}~\bibnamefont {Hebeler}},\ and\ \bibinfo {author} {\bibfnamefont {A.}~\bibnamefont {Schwenk}},\ }\bibfield  {title} {\bibinfo {title} {{Improved structure of calcium isotopes from ab initio calculations}},\ }\href {https://doi.org/10.1103/PhysRevC.111.034311} {\bibfield  {journal} {\bibinfo  {journal} {Phys. Rev. C}\ }\textbf {\bibinfo {volume} {111}},\ \bibinfo {pages} {034311} (\bibinfo {year} {2025})}\BibitemShut {NoStop}%
\bibitem [{\citenamefont {Steffen}\ \emph {et~al.}(1980)\citenamefont {Steffen}, \citenamefont {Gr{\"a}f}, \citenamefont {Gross}, \citenamefont {Meuer}, \citenamefont {Richter}, \citenamefont {Spamer}, \citenamefont {Titze},\ and\ \citenamefont {Kn{\"u}pfer}}]{Steffen1980}%
  \BibitemOpen
  \bibfield  {author} {\bibinfo {author} {\bibfnamefont {W.}~\bibnamefont {Steffen}}, \bibinfo {author} {\bibfnamefont {H.-D.}\ \bibnamefont {Gr{\"a}f}}, \bibinfo {author} {\bibfnamefont {W.}~\bibnamefont {Gross}}, \bibinfo {author} {\bibfnamefont {D.}~\bibnamefont {Meuer}}, \bibinfo {author} {\bibfnamefont {A.}~\bibnamefont {Richter}}, \bibinfo {author} {\bibfnamefont {E.}~\bibnamefont {Spamer}}, \bibinfo {author} {\bibfnamefont {O.}~\bibnamefont {Titze}},\ and\ \bibinfo {author} {\bibfnamefont {W.}~\bibnamefont {Kn{\"u}pfer}},\ }\bibfield  {title} {\bibinfo {title} {Backward-angle high-resolution inelastic electron scattering on \ensuremath{^{40, 42, 44, 48}}{{Ca}} and observation of a very strong magnetic dipole ground-state transition in \ensuremath{^{48}}{{Ca}}},\ }\href {https://doi.org/10.1016/0370-2693(80)90390-1} {\bibfield  {journal} {\bibinfo  {journal} {Phys. Lett. B}\ }\textbf {\bibinfo {volume} {95}},\ \bibinfo {pages} {23} (\bibinfo {year} {1980})}\BibitemShut {NoStop}%
\bibitem [{\citenamefont {Steffen}\ \emph {et~al.}(1983)\citenamefont {Steffen}, \citenamefont {Gr{\"a}f}, \citenamefont {Richter}, \citenamefont {H{\"a}rting}, \citenamefont {Weise}, \citenamefont {Deutschmann}, \citenamefont {Lahm},\ and\ \citenamefont {Neuhausen}}]{Steffen1983}%
  \BibitemOpen
  \bibfield  {author} {\bibinfo {author} {\bibfnamefont {W.}~\bibnamefont {Steffen}}, \bibinfo {author} {\bibfnamefont {H.-D.}\ \bibnamefont {Gr{\"a}f}}, \bibinfo {author} {\bibfnamefont {A.}~\bibnamefont {Richter}}, \bibinfo {author} {\bibfnamefont {A.}~\bibnamefont {H{\"a}rting}}, \bibinfo {author} {\bibfnamefont {W.}~\bibnamefont {Weise}}, \bibinfo {author} {\bibfnamefont {U.}~\bibnamefont {Deutschmann}}, \bibinfo {author} {\bibfnamefont {G.}~\bibnamefont {Lahm}},\ and\ \bibinfo {author} {\bibfnamefont {R.}~\bibnamefont {Neuhausen}},\ }\bibfield  {title} {\bibinfo {title} {Form factor of the {{M1}} transition to the 10.23 {{MeV}} state in \ensuremath{^{48}}{{Ca}} and the role of the {{$\Delta$}}(1232)},\ }\href {https://doi.org/10.1016/0375-9474(83)90267-1} {\bibfield  {journal} {\bibinfo  {journal} {Nucl. Phys. A}\ }\textbf {\bibinfo {volume} {404}},\ \bibinfo {pages} {413} (\bibinfo {year} {1983})}\BibitemShut {NoStop}%
\bibitem [{\citenamefont {Tompkins}\ \emph {et~al.}(2011)\citenamefont {Tompkins}, \citenamefont {Arnold}, \citenamefont {Karwowski}, \citenamefont {Rich}, \citenamefont {Sobotka},\ and\ \citenamefont {Howell}}]{Tompkins2011}%
  \BibitemOpen
  \bibfield  {author} {\bibinfo {author} {\bibfnamefont {J.~R.}\ \bibnamefont {Tompkins}}, \bibinfo {author} {\bibfnamefont {C.~W.}\ \bibnamefont {Arnold}}, \bibinfo {author} {\bibfnamefont {H.~J.}\ \bibnamefont {Karwowski}}, \bibinfo {author} {\bibfnamefont {G.~C.}\ \bibnamefont {Rich}}, \bibinfo {author} {\bibfnamefont {L.~G.}\ \bibnamefont {Sobotka}},\ and\ \bibinfo {author} {\bibfnamefont {C.~R.}\ \bibnamefont {Howell}},\ }\bibfield  {title} {\bibinfo {title} {Measurements of the \ensuremath{^{48}}{{Ca}}({$\gamma$},n) reaction},\ }\href {https://doi.org/10.1103/PhysRevC.84.044331} {\bibfield  {journal} {\bibinfo  {journal} {Phys. Rev. C}\ }\textbf {\bibinfo {volume} {84}},\ \bibinfo {pages} {044331} (\bibinfo {year} {2011})}\BibitemShut {NoStop}%
\bibitem [{\citenamefont {Birkhan}\ \emph {et~al.}(2016)\citenamefont {Birkhan}, \citenamefont {Matsubara}, \citenamefont {{von Neumann-Cosel}}, \citenamefont {Pietralla}, \citenamefont {Ponomarev}, \citenamefont {Richter}, \citenamefont {Tamii},\ and\ \citenamefont {Wambach}}]{Birkhan2016}%
  \BibitemOpen
  \bibfield  {author} {\bibinfo {author} {\bibfnamefont {J.}~\bibnamefont {Birkhan}}, \bibinfo {author} {\bibfnamefont {H.}~\bibnamefont {Matsubara}}, \bibinfo {author} {\bibfnamefont {P.}~\bibnamefont {{von Neumann-Cosel}}}, \bibinfo {author} {\bibfnamefont {N.}~\bibnamefont {Pietralla}}, \bibinfo {author} {\bibfnamefont {V.~{\relax Yu}.}\ \bibnamefont {Ponomarev}}, \bibinfo {author} {\bibfnamefont {A.}~\bibnamefont {Richter}}, \bibinfo {author} {\bibfnamefont {A.}~\bibnamefont {Tamii}},\ and\ \bibinfo {author} {\bibfnamefont {J.}~\bibnamefont {Wambach}},\ }\bibfield  {title} {\bibinfo {title} {Electromagnetic {{M1}} transition strengths from inelastic proton scattering: {{The}} cases of \ensuremath{^{48}}{{Ca}} and \ensuremath{^{208}}{{Pb}}},\ }\href {https://doi.org/10.1103/PhysRevC.93.041302} {\bibfield  {journal} {\bibinfo  {journal} {Phys. Rev. C}\ }\textbf {\bibinfo {volume} {93}},\ \bibinfo {pages} {041302} (\bibinfo {year} {2016})}\BibitemShut {NoStop}%
\bibitem [{\citenamefont {Guhr}\ \emph {et~al.}(1990)\citenamefont {Guhr}, \citenamefont {Diesener}, \citenamefont {Richter}, \citenamefont {Jager}, \citenamefont {Vries},\ and\ \citenamefont {Witt~Huberts}}]{Guhr1990}%
  \BibitemOpen
  \bibfield  {author} {\bibinfo {author} {\bibfnamefont {T.}~\bibnamefont {Guhr}}, \bibinfo {author} {\bibfnamefont {H.}~\bibnamefont {Diesener}}, \bibinfo {author} {\bibfnamefont {A.}~\bibnamefont {Richter}}, \bibinfo {author} {\bibfnamefont {C.~W.}\ \bibnamefont {Jager}}, \bibinfo {author} {\bibfnamefont {H.}~\bibnamefont {Vries}},\ and\ \bibinfo {author} {\bibfnamefont {P.~K.~A.}\ \bibnamefont {Witt~Huberts}},\ }\bibfield  {title} {\bibinfo {title} {Electroexcitation of magnetic dipole and other modes {{in \ensuremath{^{46}}Ti and \ensuremath{^{48}}Ti}}},\ }\href {https://doi.org/10.1007/BF01290617} {\bibfield  {journal} {\bibinfo  {journal} {Z. Phys. A}\ }\textbf {\bibinfo {volume} {336}},\ \bibinfo {pages} {159} (\bibinfo {year} {1990})}\BibitemShut {NoStop}%
\bibitem [{\citenamefont {Hu}\ \emph {et~al.}(2022)\citenamefont {Hu}, \citenamefont {Jiang}, \citenamefont {Miyagi}, \citenamefont {Sun}, \citenamefont {Ekstr{\"o}m}, \citenamefont {Forss{\'e}n}, \citenamefont {Hagen}, \citenamefont {Holt}, \citenamefont {Papenbrock}, \citenamefont {Stroberg},\ and\ \citenamefont {Vernon}}]{Hu2022}%
  \BibitemOpen
  \bibfield  {author} {\bibinfo {author} {\bibfnamefont {B.}~\bibnamefont {Hu}}, \bibinfo {author} {\bibfnamefont {W.}~\bibnamefont {Jiang}}, \bibinfo {author} {\bibfnamefont {T.}~\bibnamefont {Miyagi}}, \bibinfo {author} {\bibfnamefont {Z.}~\bibnamefont {Sun}}, \bibinfo {author} {\bibfnamefont {A.}~\bibnamefont {Ekstr{\"o}m}}, \bibinfo {author} {\bibfnamefont {C.}~\bibnamefont {Forss{\'e}n}}, \bibinfo {author} {\bibfnamefont {G.}~\bibnamefont {Hagen}}, \bibinfo {author} {\bibfnamefont {J.~D.}\ \bibnamefont {Holt}}, \bibinfo {author} {\bibfnamefont {T.}~\bibnamefont {Papenbrock}}, \bibinfo {author} {\bibfnamefont {S.~R.}\ \bibnamefont {Stroberg}},\ and\ \bibinfo {author} {\bibfnamefont {I.}~\bibnamefont {Vernon}},\ }\bibfield  {title} {\bibinfo {title} {Ab initio predictions link the neutron skin of \ensuremath{^{208}}{{Pb}} to nuclear forces},\ }\href {https://doi.org/10.1038/s41567-022-01715-8} {\bibfield  {journal} {\bibinfo  {journal} {Nature Phys.}\ }\textbf {\bibinfo {volume} {18}},\ \bibinfo {pages}
  {1196} (\bibinfo {year} {2022})}\BibitemShut {NoStop}%
\bibitem [{\citenamefont {Hebeler}\ \emph {et~al.}(2011)\citenamefont {Hebeler}, \citenamefont {Bogner}, \citenamefont {Furnstahl}, \citenamefont {Nogga},\ and\ \citenamefont {Schwenk}}]{Hebeler2011}%
  \BibitemOpen
  \bibfield  {author} {\bibinfo {author} {\bibfnamefont {K.}~\bibnamefont {Hebeler}}, \bibinfo {author} {\bibfnamefont {S.~K.}\ \bibnamefont {Bogner}}, \bibinfo {author} {\bibfnamefont {R.~J.}\ \bibnamefont {Furnstahl}}, \bibinfo {author} {\bibfnamefont {A.}~\bibnamefont {Nogga}},\ and\ \bibinfo {author} {\bibfnamefont {A.}~\bibnamefont {Schwenk}},\ }\bibfield  {title} {\bibinfo {title} {Improved nuclear matter calculations from chiral low-momentum interactions},\ }\href {https://doi.org/10.1103/PhysRevC.83.031301} {\bibfield  {journal} {\bibinfo  {journal} {Phys. Rev. C}\ }\textbf {\bibinfo {volume} {83}},\ \bibinfo {pages} {031301} (\bibinfo {year} {2011})}\BibitemShut {NoStop}%
\bibitem [{\citenamefont {Tsukiyama}\ \emph {et~al.}(2012)\citenamefont {Tsukiyama}, \citenamefont {Bogner},\ and\ \citenamefont {Schwenk}}]{Tsukiyama:2012sm}%
  \BibitemOpen
  \bibfield  {author} {\bibinfo {author} {\bibfnamefont {K.}~\bibnamefont {Tsukiyama}}, \bibinfo {author} {\bibfnamefont {S.~K.}\ \bibnamefont {Bogner}},\ and\ \bibinfo {author} {\bibfnamefont {A.}~\bibnamefont {Schwenk}},\ }\bibfield  {title} {\bibinfo {title} {{In-Medium Similarity Renormalization Group for Open-Shell Nuclei}},\ }\href {https://doi.org/10.1103/PhysRevC.85.061304} {\bibfield  {journal} {\bibinfo  {journal} {Phys. Rev. C}\ }\textbf {\bibinfo {volume} {85}},\ \bibinfo {pages} {061304} (\bibinfo {year} {2012})}\BibitemShut {NoStop}%
\bibitem [{\citenamefont {Morris}\ \emph {et~al.}(2015)\citenamefont {Morris}, \citenamefont {Parzuchowski},\ and\ \citenamefont {Bogner}}]{Morris:2015yna}%
  \BibitemOpen
  \bibfield  {author} {\bibinfo {author} {\bibfnamefont {T.~D.}\ \bibnamefont {Morris}}, \bibinfo {author} {\bibfnamefont {N.}~\bibnamefont {Parzuchowski}},\ and\ \bibinfo {author} {\bibfnamefont {S.~K.}\ \bibnamefont {Bogner}},\ }\bibfield  {title} {\bibinfo {title} {{Magnus expansion and in-medium similarity renormalization group}},\ }\href {https://doi.org/10.1103/PhysRevC.92.034331} {\bibfield  {journal} {\bibinfo  {journal} {Phys. Rev. C}\ }\textbf {\bibinfo {volume} {92}},\ \bibinfo {pages} {034331} (\bibinfo {year} {2015})}\BibitemShut {NoStop}%
\bibitem [{\citenamefont {Miyagi}(2023)}]{Miyagi2023}%
  \BibitemOpen
  \bibfield  {author} {\bibinfo {author} {\bibfnamefont {T.}~\bibnamefont {Miyagi}},\ }\bibfield  {title} {\bibinfo {title} {{{NuHamil}}: {{A}} numerical code to generate nuclear two- and three-body matrix elements from chiral effective field theory},\ }\href {https://doi.org/10.1140/epja/s10050-023-01039-y} {\bibfield  {journal} {\bibinfo  {journal} {Eur. Phys. J. A}\ }\textbf {\bibinfo {volume} {59}},\ \bibinfo {pages} {150} (\bibinfo {year} {2023})}\BibitemShut {NoStop}%
\bibitem [{\citenamefont {Stroberg}(2018)}]{Stroberg2018}%
  \BibitemOpen
  \bibfield  {author} {\bibinfo {author} {\bibfnamefont {S.~R.}\ \bibnamefont {Stroberg}},\ }\href {https://github.com/ragnarstroberg/imsrg} {\bibinfo {title} {{{IMSRG}}++}} (\bibinfo {year} {2018})\BibitemShut {NoStop}%
\bibitem [{\citenamefont {Shimizu}\ \emph {et~al.}(2019)\citenamefont {Shimizu}, \citenamefont {Mizusaki}, \citenamefont {Utsuno},\ and\ \citenamefont {Tsunoda}}]{Shimizu:2019xcd}%
  \BibitemOpen
  \bibfield  {author} {\bibinfo {author} {\bibfnamefont {N.}~\bibnamefont {Shimizu}}, \bibinfo {author} {\bibfnamefont {T.}~\bibnamefont {Mizusaki}}, \bibinfo {author} {\bibfnamefont {Y.}~\bibnamefont {Utsuno}},\ and\ \bibinfo {author} {\bibfnamefont {Y.}~\bibnamefont {Tsunoda}},\ }\bibfield  {title} {\bibinfo {title} {{Thick-Restart Block Lanczos Method for Large-Scale Shell-Model Calculations}},\ }\href {https://doi.org/10.1016/j.cpc.2019.06.011} {\bibfield  {journal} {\bibinfo  {journal} {Comput. Phys. Commun.}\ }\textbf {\bibinfo {volume} {244}},\ \bibinfo {pages} {372} (\bibinfo {year} {2019})}\BibitemShut {NoStop}%
\bibitem [{\citenamefont {Krebs}\ \emph {et~al.}(2019)\citenamefont {Krebs}, \citenamefont {Epelbaum},\ and\ \citenamefont {Mei{\ss}ner}}]{Krebs2019}%
  \BibitemOpen
  \bibfield  {author} {\bibinfo {author} {\bibfnamefont {H.}~\bibnamefont {Krebs}}, \bibinfo {author} {\bibfnamefont {E.}~\bibnamefont {Epelbaum}},\ and\ \bibinfo {author} {\bibfnamefont {U.-G.}\ \bibnamefont {Mei{\ss}ner}},\ }\bibfield  {title} {\bibinfo {title} {{Nuclear Electromagnetic Currents to Fourth Order in Chiral Effective Field Theory}},\ }\href {https://doi.org/10.1007/s00601-019-1500-5} {\bibfield  {journal} {\bibinfo  {journal} {Few-Body Syst.}\ }\textbf {\bibinfo {volume} {60}},\ \bibinfo {pages} {31} (\bibinfo {year} {2019})}\BibitemShut {NoStop}%
\bibitem [{\citenamefont {Krebs}(2020)}]{Krebs2020}%
  \BibitemOpen
  \bibfield  {author} {\bibinfo {author} {\bibfnamefont {H.}~\bibnamefont {Krebs}},\ }\bibfield  {title} {\bibinfo {title} {Nuclear currents in chiral effective field theory},\ }\href {https://doi.org/10.1140/epja/s10050-020-00230-9} {\bibfield  {journal} {\bibinfo  {journal} {Eur. Phys. J. A}\ }\textbf {\bibinfo {volume} {56}},\ \bibinfo {pages} {234} (\bibinfo {year} {2020})}\BibitemShut {NoStop}%
\bibitem [{\citenamefont {Kamuntavi{\v{c}}ius}\ \emph {et~al.}(2001)\citenamefont {Kamuntavi{\v{c}}ius}, \citenamefont {Kalinauskas}, \citenamefont {Barrett}, \citenamefont {Mickevi{\v{c}}ius},\ and\ \citenamefont {Germanas}}]{Kamuntavicius2001}%
  \BibitemOpen
  \bibfield  {author} {\bibinfo {author} {\bibfnamefont {G.}~\bibnamefont {Kamuntavi{\v{c}}ius}}, \bibinfo {author} {\bibfnamefont {R.}~\bibnamefont {Kalinauskas}}, \bibinfo {author} {\bibfnamefont {B.}~\bibnamefont {Barrett}}, \bibinfo {author} {\bibfnamefont {S.}~\bibnamefont {Mickevi{\v{c}}ius}},\ and\ \bibinfo {author} {\bibfnamefont {D.}~\bibnamefont {Germanas}},\ }\bibfield  {title} {\bibinfo {title} {{The general harmonic-oscillator brackets: compact expression, symmetries, sums and Fortran code}},\ }\href {https://doi.org/10.1016/S0375-9474(01)01101-0} {\bibfield  {journal} {\bibinfo  {journal} {Nucl. Phys. A}\ }\textbf {\bibinfo {volume} {695}},\ \bibinfo {pages} {191} (\bibinfo {year} {2001})}\BibitemShut {NoStop}%
\bibitem [{\citenamefont {Virtanen}\ \emph {et~al.}(2020)\citenamefont {Virtanen} \emph {et~al.}}]{Virtanen2020}%
  \BibitemOpen
  \bibfield  {author} {\bibinfo {author} {\bibfnamefont {P.}~\bibnamefont {Virtanen}} \emph {et~al.},\ }\bibfield  {title} {\bibinfo {title} {{{SciPy}} 1.0: Fundamental algorithms for scientific computing in {{Python}}},\ }\href {https://doi.org/10.1038/s41592-019-0686-2} {\bibfield  {journal} {\bibinfo  {journal} {Nature Meth.}\ }\textbf {\bibinfo {volume} {17}},\ \bibinfo {pages} {261} (\bibinfo {year} {2020})}\BibitemShut {NoStop}%
\bibitem [{\citenamefont {Grewe}\ \emph {et~al.}(2007)\citenamefont {Grewe}, \citenamefont {Frekers}, \citenamefont {Rakers}, \citenamefont {Adachi}, \citenamefont {B{\"a}umer}, \citenamefont {Botha}, \citenamefont {Dohmann}, \citenamefont {Fujita}, \citenamefont {Fujita}, \citenamefont {Hatanaka}, \citenamefont {Nakanishi}, \citenamefont {Negret}, \citenamefont {Neveling}, \citenamefont {Popescu}, \citenamefont {Sakemi}, \citenamefont {Shimbara}, \citenamefont {Shimizu}, \citenamefont {Smit}, \citenamefont {Tameshige}, \citenamefont {Tamii}, \citenamefont {Thies}, \citenamefont {von Brentano}, \citenamefont {Yosoi},\ and\ \citenamefont {Zegers}}]{Grewe2007}%
  \BibitemOpen
  \bibfield  {author} {\bibinfo {author} {\bibfnamefont {E.-W.}\ \bibnamefont {Grewe}}, \bibinfo {author} {\bibfnamefont {D.}~\bibnamefont {Frekers}}, \bibinfo {author} {\bibfnamefont {S.}~\bibnamefont {Rakers}}, \bibinfo {author} {\bibfnamefont {T.}~\bibnamefont {Adachi}}, \bibinfo {author} {\bibfnamefont {C.}~\bibnamefont {B{\"a}umer}}, \bibinfo {author} {\bibfnamefont {N.~T.}\ \bibnamefont {Botha}}, \bibinfo {author} {\bibfnamefont {H.}~\bibnamefont {Dohmann}}, \bibinfo {author} {\bibfnamefont {H.}~\bibnamefont {Fujita}}, \bibinfo {author} {\bibfnamefont {Y.}~\bibnamefont {Fujita}}, \bibinfo {author} {\bibfnamefont {K.}~\bibnamefont {Hatanaka}}, \bibinfo {author} {\bibfnamefont {K.}~\bibnamefont {Nakanishi}}, \bibinfo {author} {\bibfnamefont {A.}~\bibnamefont {Negret}}, \bibinfo {author} {\bibfnamefont {R.}~\bibnamefont {Neveling}}, \bibinfo {author} {\bibfnamefont {L.}~\bibnamefont {Popescu}}, \bibinfo {author} {\bibfnamefont {Y.}~\bibnamefont {Sakemi}}, \bibinfo {author} {\bibfnamefont {Y.}~\bibnamefont
  {Shimbara}}, \bibinfo {author} {\bibfnamefont {Y.}~\bibnamefont {Shimizu}}, \bibinfo {author} {\bibfnamefont {F.~D.}\ \bibnamefont {Smit}}, \bibinfo {author} {\bibfnamefont {Y.}~\bibnamefont {Tameshige}}, \bibinfo {author} {\bibfnamefont {A.}~\bibnamefont {Tamii}}, \bibinfo {author} {\bibfnamefont {J.}~\bibnamefont {Thies}}, \bibinfo {author} {\bibfnamefont {P.}~\bibnamefont {von Brentano}}, \bibinfo {author} {\bibfnamefont {M.}~\bibnamefont {Yosoi}},\ and\ \bibinfo {author} {\bibfnamefont {R.~G.~T.}\ \bibnamefont {Zegers}},\ }\bibfield  {title} {\bibinfo {title} {(\ensuremath{^3}{{He}},$t$) reaction on the double {$\beta$} decay nucleus \ensuremath{^{48}}{{Ca}} and the importance of nuclear matrix elements},\ }\href {https://doi.org/10.1103/PhysRevC.76.054307} {\bibfield  {journal} {\bibinfo  {journal} {Phys. Rev. C}\ }\textbf {\bibinfo {volume} {76}},\ \bibinfo {pages} {054307} (\bibinfo {year} {2007})}\BibitemShut {NoStop}%
\bibitem [{\citenamefont {{Information extracted from the NuDat database}}()}]{Nndc}%
  \BibitemOpen
  \bibfield  {author} {\bibinfo {author} {\bibnamefont {{Information extracted from the NuDat database}}},\ }\href@noop {} {}\bibinfo {howpublished} {\url{https://www.nndc.bnl.gov/nudat3/}}\BibitemShut {NoStop}%
\bibitem [{\citenamefont {Richter}(1990)}]{Richter1990}%
  \BibitemOpen
  \bibfield  {author} {\bibinfo {author} {\bibfnamefont {A.}~\bibnamefont {Richter}},\ }\bibfield  {title} {\bibinfo {title} {Shell model and magnetic dipole modes in deformed nuclei},\ }\href {https://doi.org/10.1016/0375-9474(90)90571-3} {\bibfield  {journal} {\bibinfo  {journal} {Nucl. Phys. A}\ }\textbf {\bibinfo {volume} {507}},\ \bibinfo {pages} {99} (\bibinfo {year} {1990})}\BibitemShut {NoStop}%
\bibitem [{\citenamefont {Brase}\ \emph {et~al.}()\citenamefont {Brase}, \citenamefont {Miyagi}, \citenamefont {Men{\'e}ndez},\ and\ \citenamefont {Schwenk}}]{Zenodo}%
  \BibitemOpen
  \bibfield  {author} {\bibinfo {author} {\bibfnamefont {C.}~\bibnamefont {Brase}}, \bibinfo {author} {\bibfnamefont {T.}~\bibnamefont {Miyagi}}, \bibinfo {author} {\bibfnamefont {J.}~\bibnamefont {Men{\'e}ndez}},\ and\ \bibinfo {author} {\bibfnamefont {A.}~\bibnamefont {Schwenk}},\ }\bibfield  {title} {\bibinfo {title} {{Data: Two-body currents at finite momentum transfer and applications to M1 transitions,}}\ }\href {https://doi.org/10.5281/zenodo.17160362} {10.5281/zenodo.17160362}\BibitemShut {NoStop}%
\bibitem [{\citenamefont {Walecka}(1995)}]{Walecka:1995mi}%
  \BibitemOpen
  \bibfield  {author} {\bibinfo {author} {\bibfnamefont {J.~D.}\ \bibnamefont {Walecka}},\ }\bibfield  {title} {\bibinfo {title} {{Theoretical nuclear and subnuclear physics}},\ }\href {https://doi.org/10.1142/5500} {\bibfield  {journal} {\bibinfo  {journal} {Oxford Stud. Nucl. Phys.}\ }\textbf {\bibinfo {volume} {16}},\ \bibinfo {pages} {1} (\bibinfo {year} {1995})}\BibitemShut {NoStop}%
\end{thebibliography}%

\end{document}